\documentclass{WileyMSP-template}
{}
{}
{}
\usepackage{amsmath}
\usepackage{subcaption}
\usepackage{graphicx}
\usepackage[utf8]{inputenc}
\usepackage[T1]{fontenc}
\usepackage{textcomp}
\usepackage{appendix}
\usepackage{url}
\usepackage{hyperref}

\begin{document}

\pagestyle{fancy}
\rhead{\includegraphics[width=2.5cm]{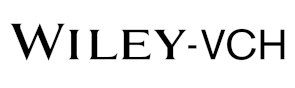}}

\title{A Modelling study of Electron transport in GaN/AlGaN superlattices using Monte Carlo simulation}

\maketitle


\author{Mengxun Bai}
\author{Judy Rorison}


\dedication{}

\begin{affiliations}
Mengxun Bai and Judy Rorison\\
Electrical and Electronic Engineering, University of Bristol, Bristol, UK\\
Email Addresses:mb18200@bristol.ac.uk judy.rorison@bristol.ac.uk

\end{affiliations}



\begin{abstract}

Electron transport in GaN/Al$_x$Ga$_{1-x}$N superlattices is investigated using a single particle Monte Carlo approach. To establish the band structure required GaN, AlN and their ternary alloy are investigated using a single electron Monte Carlo approach and a 3-band approximation to the full band structure. The interplay of the inter-valley scattering and electron-longitudinal optical polar phonon scattering in determining electron velocity and velocity overshoot is examined for the binaries and their alloy. However, it is observed that both scattering processes cause velocity overshoot with their interplay leading to the magnitude of the overshoot and the value of the electric field required to achieve it. A single non-parabolic band approximation is found to be acceptable to be used in the superlattice modeling because the energy width of the miniband is such that the kinetic energy of the electrons in it would not be sufficient to suffer inter-valley scattering. We use a Schrödinger wave equation coupled to a Poisson solver to self-consistently calculate the energy band structure of the superlattice using the single band approximation for the materials, determine the Fermi energy and the superlattice miniband energy position and its energy width.  We then analyze the miniband band structure and determine the effective masses for the superlattice miniband in the superlattice direction which will determine the electron mobility in that direction. Then the single particle Monte Carlo method is applied to investigate electron transport in the miniband where we find that for low Al concentration in the barrier and short periods electron velocity, very similar to that in bulk GaN can be obtained and observe that velocity overshoot can occur, purely due to electron-LO phonon scattering and non-parabolicity in the single band. This modeling approach provides a fast and convenient method to investigate high-field electron transport in n-doped GaN/Al$_x$Ga$_{1-x}$N superlattices and should be suitable for use in device design.

\end{abstract}


\section{Introduction}
GaN, AlN and their related alloy Al$_x$Ga$_{1-x}$N are considered important materials for electronic and optoelectronic device applications as they span a wide range of energy gaps and offer large breakdown fields and high thermal conductivity. These properties lead to higher output power and frequency performance of electronic devices made from these materials[1].  In optoelectronics, tuneable energy bandgaps in the blue and UV regions are beneficial for novel optoelectronic device applications[2]. Related quaternary nitride alloys have proven promising in optoelectronics for applications in blue-green and blue-violet light-emitting diodes(LEDs), laser diodes(LDs) and photodetectors[3].  High electron mobility transistors(HEMTs) based on the wurtzite phase of AlGaN/GaN heterostructures have been extensively studied[4][5]. To fully exploit these material systems it is necessary to understand electron transport, particularly high electric field transport. Electron transport in bulk III-V nitrides has been studied over the years experimentally[6] and theoretically[1][7-10] but reproducible experimental results are relatively recent. The magnitude of velocity overshoot in GaN-based materials has been an area of disagreement and focus. 

A suitable approach for treating electronic transport including multiple scattering processes is Monte Carlo simulation[1][6][7]. For establishing equilibrium conditions single electron Monte Carlo is sufficient while for exploiting non-equilibrium electronic distributions ensemble Monte Carlo is required[11]. Monte Carlo simulation, particularly for high electric fields, relies on a band structure that is accurate to high energy, which is numerically intensive and time-consuming. For device modeling purposes, simplification of the band structure is required.  A 3-band model including the two lowest higher bands has been found to give good agreement of velocity field characteristics with fuller band structure calculations for GaN and AlN so we follow this approach[1][9]. The drift velocity as a function of the electric(F) field, average electron energy as a function of the F field, and occupation of the three valleys as a function of the F field are examined for the two binaries, and the behavior is analyzed in terms of the physics of the materials. In particular, the analysis focuses specifically on understanding how the scattering processes interact with one another and collectively influence electron behavior. For the ternary alloy Al$_x$Ga$_{1-x}$N, there is still a lack of results describing the full band structure and material parameters. In this paper, we develop a 3-band model for AlGaN and investigate how the characteristics can be tailored through alloying.

Electron transport characteristics in superlattice structures, chiefly in GaAs/AlGaAs systems[12] have received much attention, and more recently GaN/AlGaN superlattices have started to be studied[13][14]. The idea of using superlattices to tailor the band structure and exploit transport in a miniband in the vertical direction could aid devices that require electrons and holes in an active region such as lasers or which require electron transport[14][15]. P-doped superlattices in GaN/AlGaN have been investigated for decreasing hole resistivity in various devices, particularly LEDs and lasers[15]. However, there have been fewer studies on n-doped superlattices in GaN/AlGaN[13]. Superlattices allow the energy band to be tailored in position and width, introducing added design control in electronic and optoelectronic devices. Therefore although they are not required to thermalise n-dopants in GaN or low AlGaN materials they might have other advantages for electron transport. 

This paper is organized as follows. section II introduces the 3-band model to the full band structure for the bulk materials and the scattering mechanisms, comparing different bulk materials. We investigate the importance of the upper valleys and their interaction with the e-LO phonon and IV scattering rates in determining velocity-field characteristics. In Section III, we use a Schrödinger wave equation and Poisson solver coupled self-consistently to evaluate the conduction band edges and miniband properties for GaN/Al$_x$Ga$_{1-x}$N superlattices(SLs), using as input a one band non-parabolic model for the conduction band validated in the previous section. In the second part of Section III, a single particle Monte Carlo is used to simulate electron transport within the superlattice miniband. The scattering mechanisms in the superlattice miniband are investigated and the electron transport is modeled. Monte Carlo simulation results are presented for a range of GaN/AlGaN superlattices. The velocity field characteristics are presented for these superlattices and discussed. Finally, we summarize the work in Section IV and discuss the implications for device design.

\section{Modeling for bulk material}
\subsection{Band structure for GaN, AlN and AlGaN}

The single-electron Monte Carlo simulation method was applied to study the electron transport based on the generic 3-valley electron band structure shown in Fig.1. The Kane mode was applied to deal with the non-parabolicity of the lowest valley, which can only be assumed to be parabolic for low F fields[1]. The following equations show the relationship between wave vector and energy in this model. In eq.1, $k$ is the magnitude of the crystal momentum, $E$ is the electron energy, $E_g$ is the energy gap, $m^*_e$ is the effective mass of the electron, $m_0$ is the mass of the electron in free space, $h$ is the symbol for Plank's constant, and $\alpha$ is the non-parabolicity factor.

\begin{figure}[!t]
\graphicspath{ {./figure_300/} }
\centering{\includegraphics[width=0.5\textwidth]{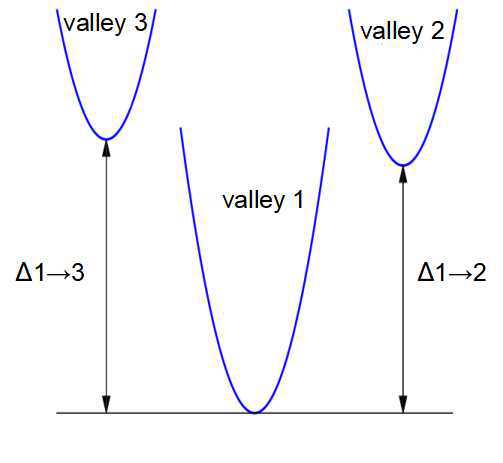}}
\caption{The 3-valley model is used to represent the conduction band structure of nitrides.}
\end{figure}

\begin{subequations}
\begin{align}
\frac{h^2k^2}{2m^*_e} & =E\left( {1 +E\alpha } \right)\\
\alpha & =\frac{1}{E_g}\left( {1 -E\frac{m^*_e}{m_0} } \right)^2
\end{align}
\label{eq1}
\end{subequations}

 Modeling of the energy band structure of the GaN and AlN has been widely studied[1][7][9]. Various methods are used to evaluate the full energy band structure or the multi-valley model for simulation[7-10][16]. The multi-band model most commonly used is the 3-valley model but the 2-valley model is also used both of which enable faster calculation than using a full-band structure approach[9]. In this paper, we use a 3-valley model for bulk materials.

The band structure parameters used for wurtzite bulk GaN and AlN are shown in Table 1[1][17][18]. Research on multi-component alloys depends on the alloy atoms' physical placement, which can be modeled for each orientation and then averaged[19][20]. It can be difficult to get good reproducible experimental values to test the modeling making the choice of parameters open to some choice. We have followed the usual approach of including the bowing parameter $b$ for the calculation of the energy gap of the ternary alloys Al$_x$Ga$_{1-x}$N as shown in eq.2 [19] where A and B can be Ga or Al. All bowing factors are selected from the reference[19][20]. Other material parameters are calculated by linear extrapolation, based on the composition.

\begin{table}
 \caption{The conduction valley parameter selections correspond to bulk wurtzite GaN and AlN}
 \centering
  \begin{tabular}[htbp]{@{}lllll@{}}
    \hline
    Material & Parameters & Valley1 & Valley2 & Valley3 \\
    \hline
    GaN     &Valley location   &$\Gamma_1^c$  &$\Gamma_3^c$  &L-M \\
            &Valley degeneracy &1             &1             &6 \\
            &$\Gamma m^*_e$(isotropic)    &0.2$m_0$      &$m_0$         &$m_0$ \\
            &Energy gap(eV)    &3.39          &5.29          &5.49 \\
            &Non-parabolicity ($eV^{-1}$)        &0.189 &0  &0\\
    \hline
    AlN     &Valley location   &$\Gamma_1^c$  &L-M           &K \\
            &Valley degeneracy &1             &6             &2 \\
            &$\Gamma m^*_e$(z)    &0.48$m_0$     &$m_0$         &$m_0$\\
            &$\Gamma m^*_e$(xy)  &0.3$m_0$       &$m_0$         &$m_0$\\
            &Energy gap(eV)    &6.2           &6.9           &7.2 \\
            &Non-parabolicity ($eV^{-1}$)  &0.044 &0  &0\\
    \hline
  \end{tabular}
\end{table}

\begin{equation}
\begin{split}
E_{gA_xB_{1-x}N}(x)= xE_{gAN}+( 1-x)E_{gBN}-b_{ABN}x( 1-x)\\
\end{split}
\label{eq2}
\end{equation}

After determining the energy gap of the central valley, it is still necessary to determine the energy gaps of satellite valleys in this 3-valley model. The positions of the lowest satellite valleys are not the same for different binary alloys, as shown in Table 1. However, it can be seen that the central valley of all binary alloys occurs at $\Gamma_1^c$, and the secondary minima of the band structures are located at L-M, K, and $\Gamma_3^c$. On the basis of the band structure parameters calculated by Fang et al.[8] and Goano et al.[19], the bowing parameters at M, K, A, L and $\Gamma$ of the second and third conduction bands reported by Goano et al[19], the energy gap ranges of Al$_x$Ga$_{1-x}$N can be calculated as shown in Fig.2. Fig.2 shows the energy gap of $\Gamma_1^c$, L-M, K and $\Gamma_3^c$ minima as a function of the concentration fraction of Al. We need to select the three valleys with the lowest energy gap from the four valleys to form the model, so the solid line in the figure represents the selected valleys which change when band crossings occur as shown in the figures. When the Al concentration factor is larger than 0.38, the K valley is lower than $\Gamma_3^c$ and becomes one of the selected valleys. The 3-valley model at this time consists of $\Gamma_1^c$, L-M, and K.

\begin{figure}[!t]
\graphicspath{ {./figure_300/} }
\centering{\includegraphics[width=0.5\textwidth]{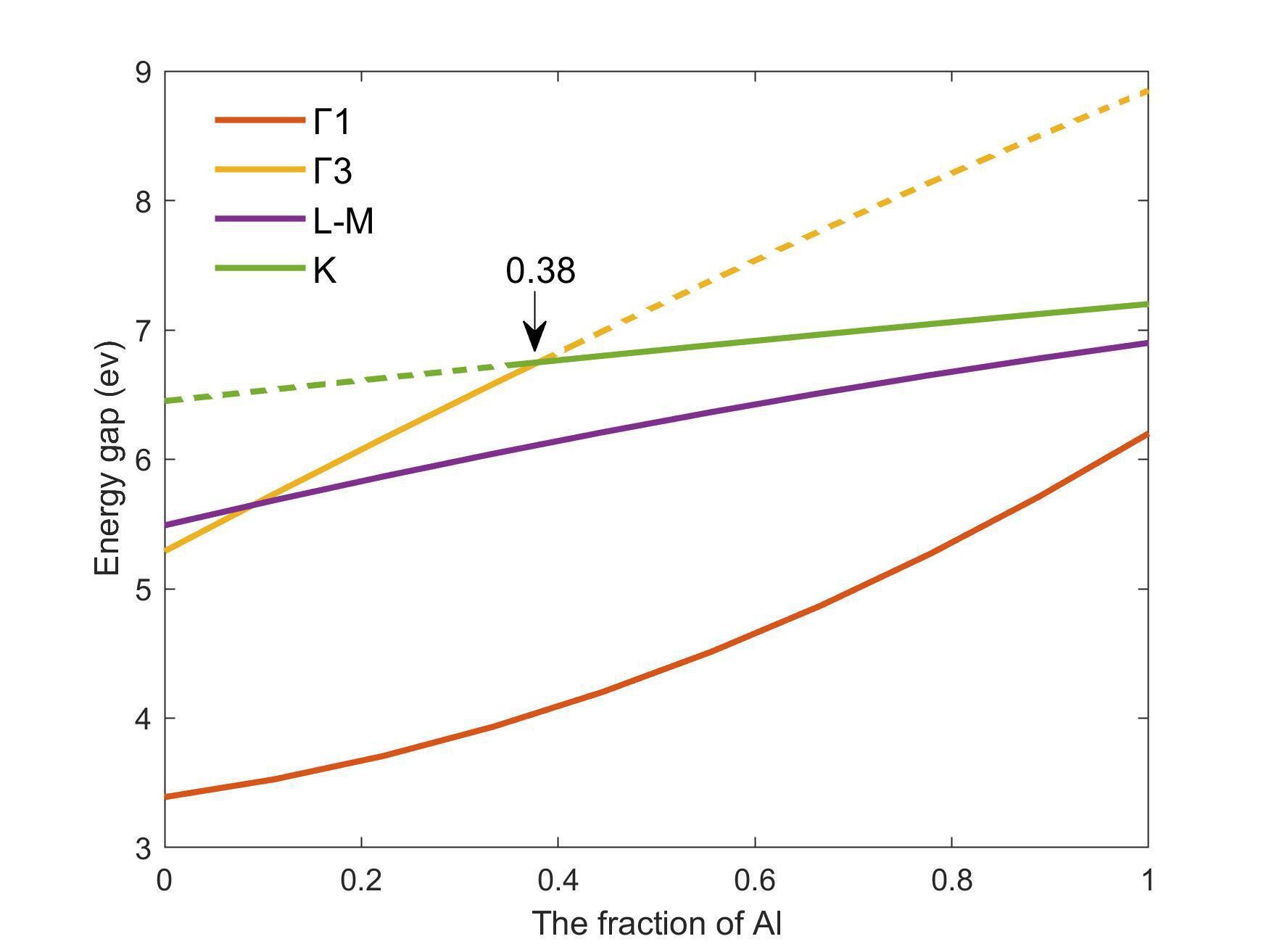}}
\caption{ The energy gap of the four lowest valleys in Al$_x$Ga$_{1-x}$N as a function of x shows the three lowest valleys to be used in the three-valley model as a function of x.}
\end{figure}

\subsection{Monte Carlo results of GaN, AlN, AlGaN}

The scattering mechanisms considered are longitudinal polar optical phonon(LPOP) scattering, ionized impurity scattering, acoustic deformation scattering, piezoelectric scattering, and inter-valley (IV) scattering. The derivations of all scattering rates come from reference[21][22]. The scattering parameters for the binaries are taken from[1][18][23-25]. The scattering parameters for the ternary alloys are determined using linear interpolation. Fig.3 shows the various scattering rates of bulk wurtzite GaN for n-doping of $10^{17}cm^{-3}$.  For the scattering mechanism with both absorption and emission scattering rate, only the emission scattering rates are shown here because they are significantly larger than the absorption scattering rates. At low electron energies, the LPOP scattering rate is the dominant scattering mechanism while for higher electron energy,  IV scattering becomes important. Fig.4 shows the scattering rate of bulk wurtzite AlN for n-doping of $10^{17}cm^{-3}$. Combined with Fig.3, we can intuitively compare the dominant scattering rate of GaN and AlN nitrides.

\begin{figure}
\centering
\begin{subfigure}{0.49\textwidth}
    \centering
    \includegraphics[width=\textwidth]{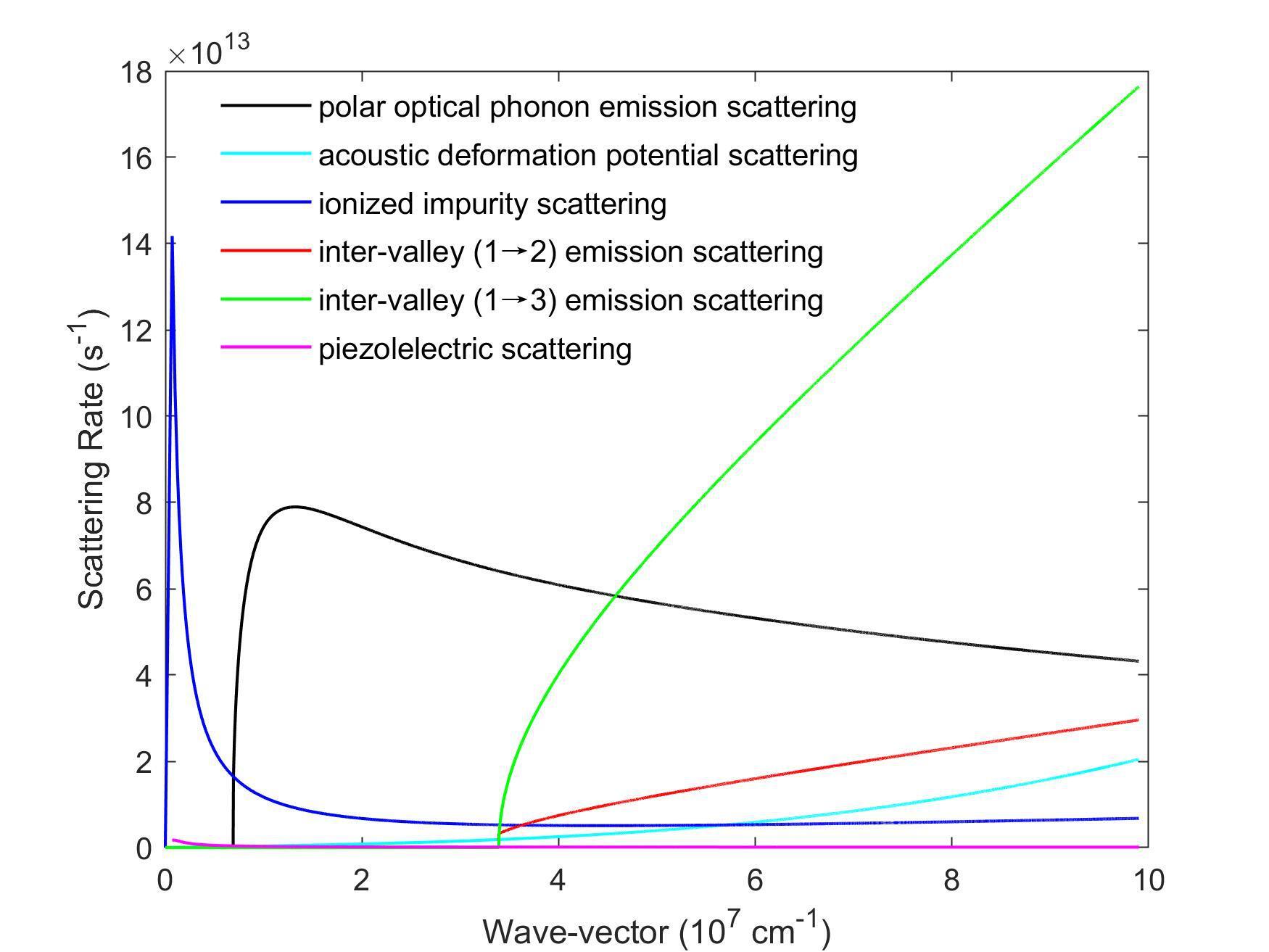}
    \caption{}
    \label{fig: first}
\end{subfigure}
\hfill
\begin{subfigure}{0.49\textwidth}
    \centering
    \includegraphics[width=\textwidth]{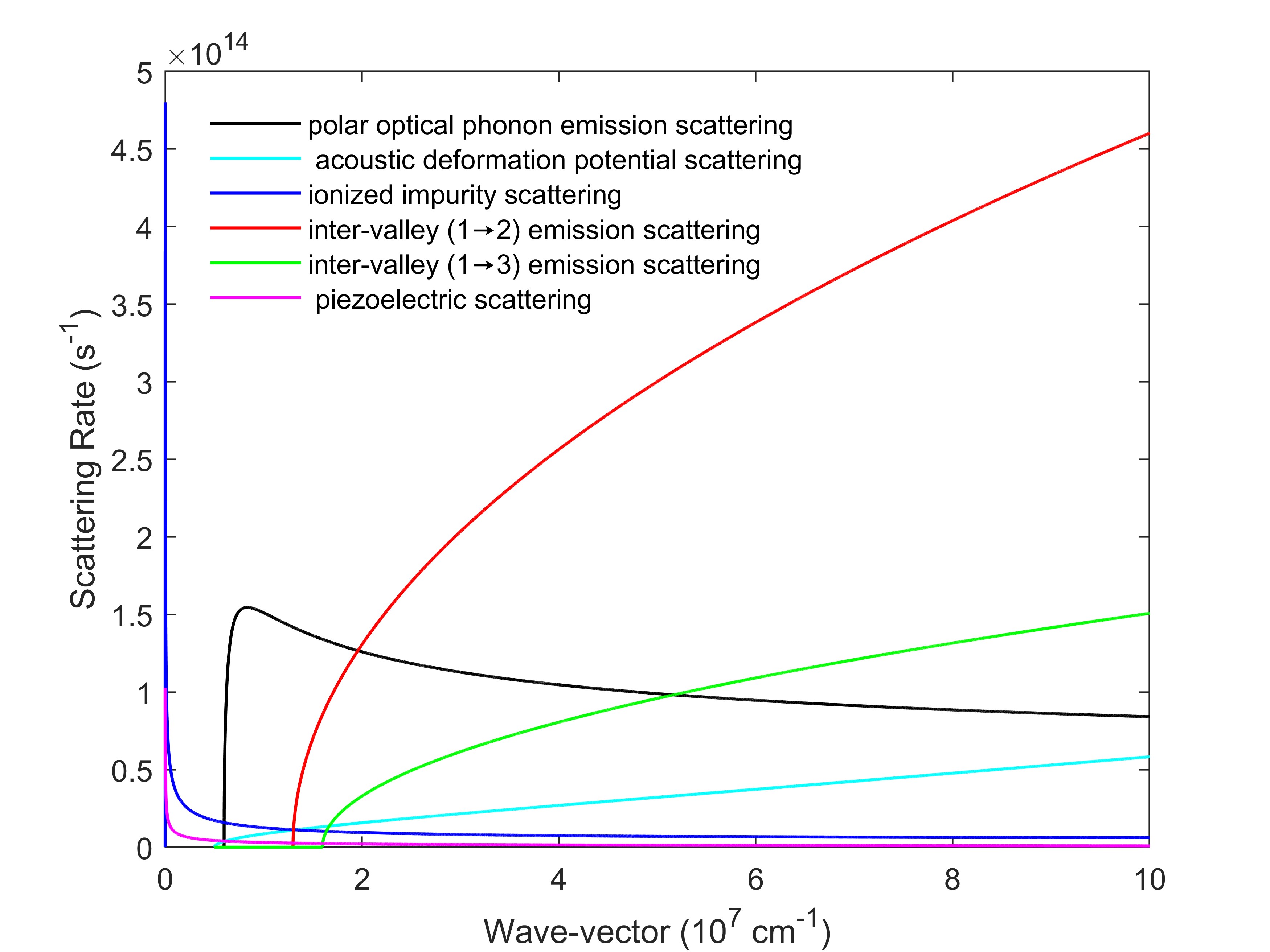}
    \caption{}
    \label{fig: second}
\end{subfigure}
\caption{(a), (b) shows the scattering rates of bulk wurtzite GaN and AlN for n-doping of $10^{17}cm^{-3}$ at T=300K. The scattering mechanisms are: ionized impurity, longitudinal polar optical phonon emission, inter-valley (1 → 3) emission, inter-valley (1 → 2) emission, acoustic deformation potential and piezoelectric.}
\label{fig: figures}
\end{figure}

Employing the 3-valley band structure model and the scattering rates shown before, the steady-state electron transport characteristics of bulk wurtzite GaN with an n-doping concentration of $10^{17}cm^{-3}$ at 300K are obtained as shown in Fig.4. Fig.4a shows the electron velocity as a function of the F field where all the scattering rates are included (solid line), where the IV scattering is omitted (dashed line) and for further comparison where the IV scattering is included but the LPOP scattering is omitted(dotted line). Although these comparisons are un-physical they allow us to investigate the interplay of the LPOP scattering and the IV scattering. Interestingly in all of these cases, a velocity overshoot is observed.

For the case where all the scattering rates are included the electron drift velocity increases rapidly with increasing F field up to an F field of about 100kV/cm. A peak velocity of $2.85\times10^7$cm/s is obtained at an F field of 166kV/cm. As the F field is increased further the velocity decreases and levels out to around $1.5\times10^7$cm/s at F fields of 600kV/cm. Therefore wurtzite GaN is observed to exhibit a velocity overshoot similar to GaAs[26][27]. We observe that neglecting the IV scattering to the upper valleys increases the magnitude of the velocity peak overshoot and moves this peak position to a higher F field value. In addition, at higher F fields the electron velocity is higher than that including all scattering rates due in part to the lower effective electron masses of the central valley. We observe that including the IV scattering but neglecting the LPOP scattering produces a very different intermediate F field velocity and produces a very high magnitude velocity overshoot peak at a relatively low F field in comparison to the curve including all the scattering rates. This shows that the interplay between the LPOP scattering and the IV scattering is dominant in determining electron velocity at intermediate and high F fields but that the former scattering rate is most important with IV scattering modifying the velocity overshoot peak value and high field velocity.

Fig.4b shows the electron occupancy of the three valleys as a function of the F field for the full scattering rate case(For the case with no IV scattering the electrons are always confined to the central valley). For F fields below 100kV/cm, the electrons occupy the central valley(valley 1) while for increasing F field, IV scattering moves the electrons to occupy the higher valleys. Valley 2 is lower in energy than valley 3 so electrons enter valley 2 at a slightly lower field but valley 3 has higher degeneracy so its occupancy increases more rapidly with increasing field but its occupancy stays below valley 2. At 200kV/cm more electrons are in the higher valleys than in the central valley. At high F fields, almost all the electrons are in the higher valleys which have higher electron effective masses than the central valley and therefore lower velocities which causes the lower velocities at the higher F fields shown.

Fig.4c shows the average electron energy as a function of the F field where the average electron energy increases sharply between 100kV/cm and 200kV/cm and then levels off to 2.1eV above an F field of 500kV/cm for the case with all the scattering rates included(solid-line). This reflects the scattering into the higher valleys providing an increase in the total density of states at the same energy enabling enhanced scattering to more states at the same energy. The average electron energy omitting IV scattering (dashed-line) has the same low and intermediate field behaviour but the average electron energy continues to increase with increasing F field reflecting the absence of the density of states scattering associated with the higher valleys.

\begin{figure}
\centering
\begin{subfigure}{0.49\textwidth}
    \centering
    \includegraphics[width=\textwidth]{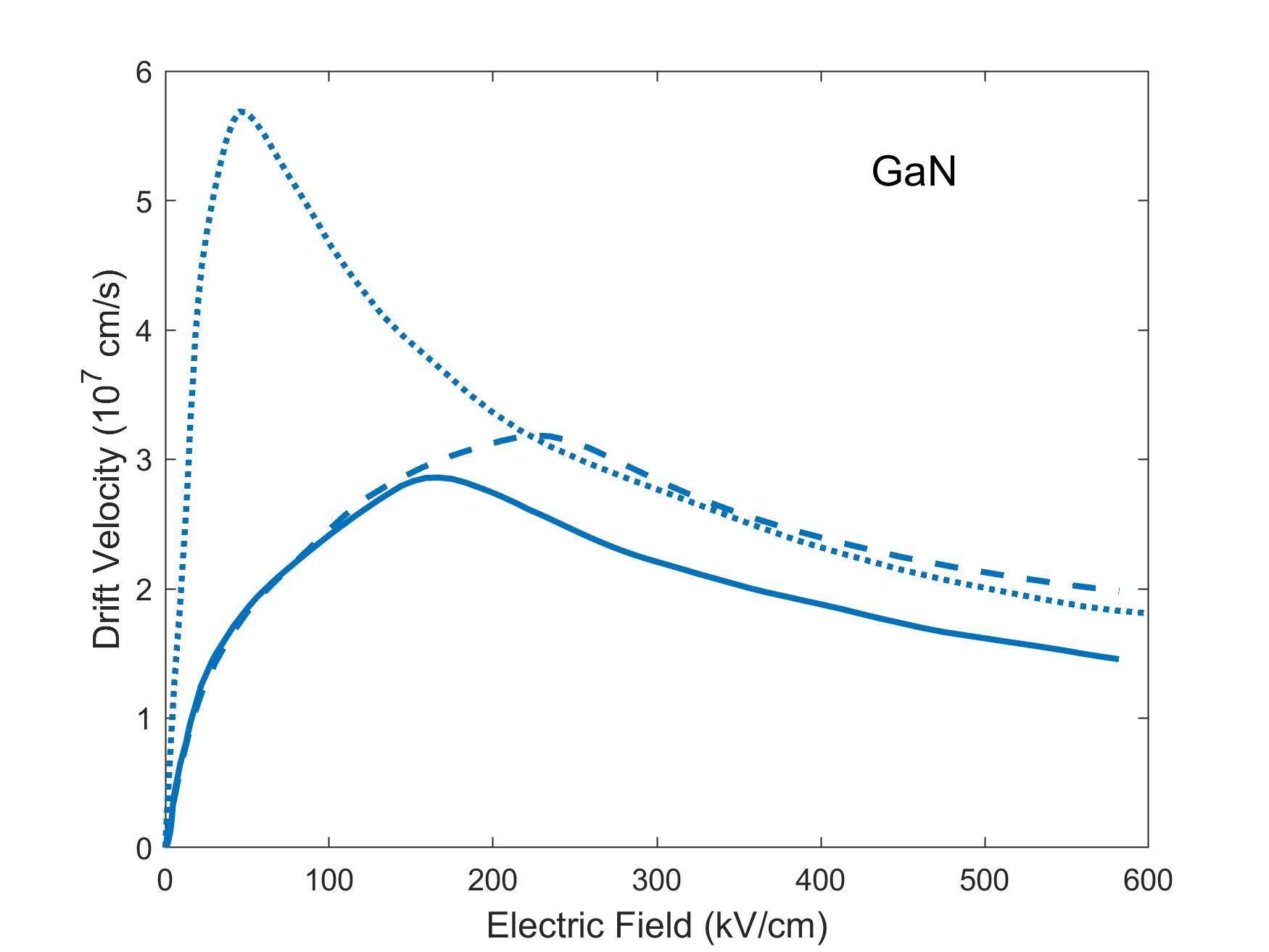}
    \caption{}
    \label{fig: first}
\end{subfigure}
\hfill
\begin{subfigure}{0.49\textwidth}
    \centering
    \includegraphics[width=\textwidth]{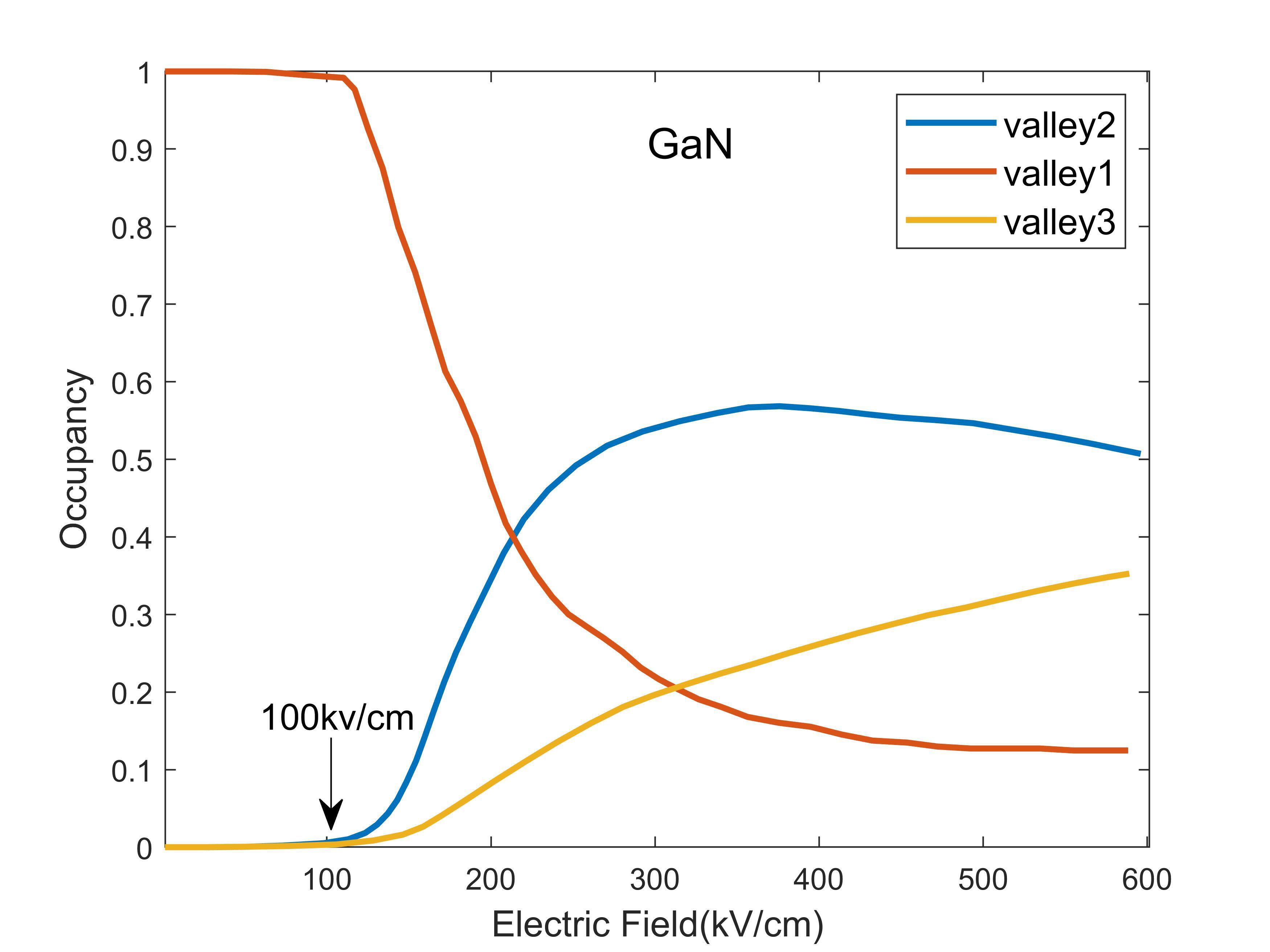}
    \caption{}
    \label{fig: second}
\end{subfigure}
\hfill
\begin{subfigure}{0.49\textwidth}
    \centering
    \includegraphics[width=\textwidth]{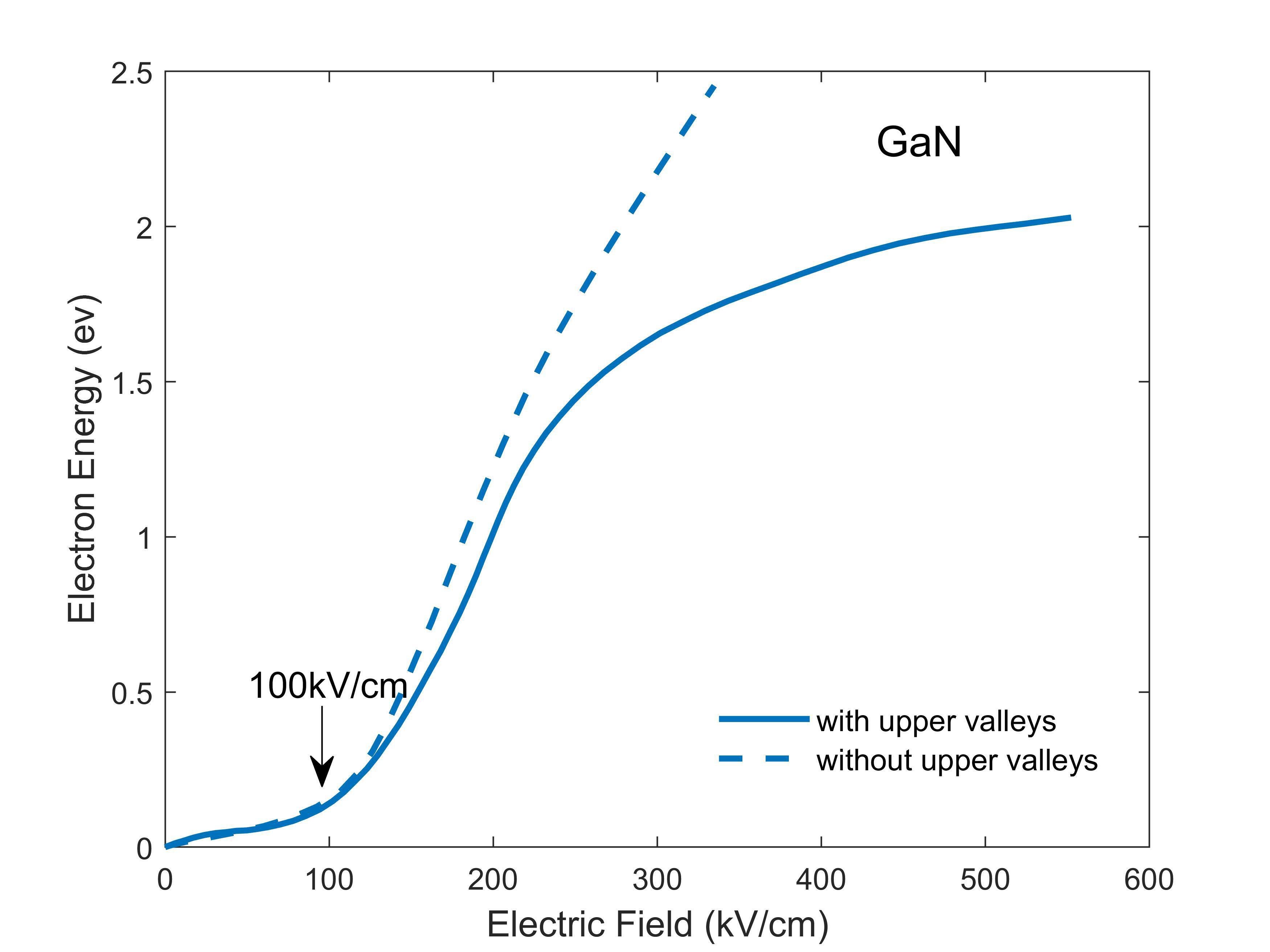}
    \caption{}
    \label{fig: second}
\end{subfigure}
\caption{(a) shows the velocity as a function of the electric field associated with bulk wurtzite GaN with n=$10^{17}cm^{-3}$ at 300K. The solid line is all considered in the 3-valley model, the dashed line excludes scattering to the upper valleys, and the dotted line excludes longitudinal polar optical phonon scattering but includes inter-valley scattering. (b) shows the valley occupancy as a function of the electric field associated with bulk wurtzite GaN. (c) shows the average electron energy as a function of the electric field with and without upper valleys associated with bulk wurtzite GaN.} 
\label{fig: figures}
\end{figure}

The electron velocity as a function of F field results of our simulations including all the scattering rates are in general agreement with previously reported results which is a good check of the model and can be used to confirm the effectiveness of the 3-valley model against simulations published using full energy band structure[1][7-10]. The experimental result shows a lower velocity as a function of the F field but shows a similar shape. It is expected that crystal imperfections and defects will lower the electron velocity[5].

The electron transport properties of AlN are also calculated and provide an interesting comparison to GaN. Fig.5a shows the electron velocity as a function of the F field for AlN with all the scattering rates included (solid line), without IV scattering (dashed line) and including IV scattering but neglecting LPOP scattering (dotted line). AlN requires the largest F field range to reach the velocity overshoot of $1.65\times10^7$cm/s which occurs at F=480kV/cm, and the saturation velocity, close to $1.45\times10^7$cm/s is not obtained until the F field strength increases over 900kV/cm. AlN exhibits a smaller peak velocity overshoot effect compared to GaN and the LPOP scattering is dominant in determining the electron velocity with only minor deviation from the curve omitting IV scattering for an F field greater than 300kV/cm. This small velocity overshoot is due to the very large energy separation between the central and upper valleys, by the relatively small difference in electron effective masses between the central valley (0.48$m_0$) and the two upper valleys (1$m_0$) and the relatively high effective masses of the central valley requiring a large F field to accelerate the electrons to high energy within the band.  Fig.3 shows the large value of LPOP scattering in AlN compared to GaN which acts to suppress the electron velocity. The velocity omitting LPOP scattering and including IV scattering shows a weak velocity overshoot at a smaller F field reflecting the similarity of the effective masses in the three valleys and the strength of the LPOP scattering in AlN.

Fig.5b shows the electron occupation of the valleys as a function of the F field showing that F fields of over 300kV/cm are required to significantly populate the upper valleys. Valley 2 is at significantly lower energy than valley 3 and has higher degeneracy so the electrons enter valley 2 first and its occupation grows with increasing field while the population in valley 3 remains small. In GaN the total population of the upper valleys at high field is larger than in GaN. Fig.5c shows the average electron energy as a function of the F field with (solid line) and without IV scattering(dashed line) reflecting that high F fields are required to populate the upper valleys. A 2-valley model would be sufficient to model AlN.

\begin{figure}
\centering
\begin{subfigure}{0.49\textwidth}
    \centering
    \includegraphics[width=\textwidth]{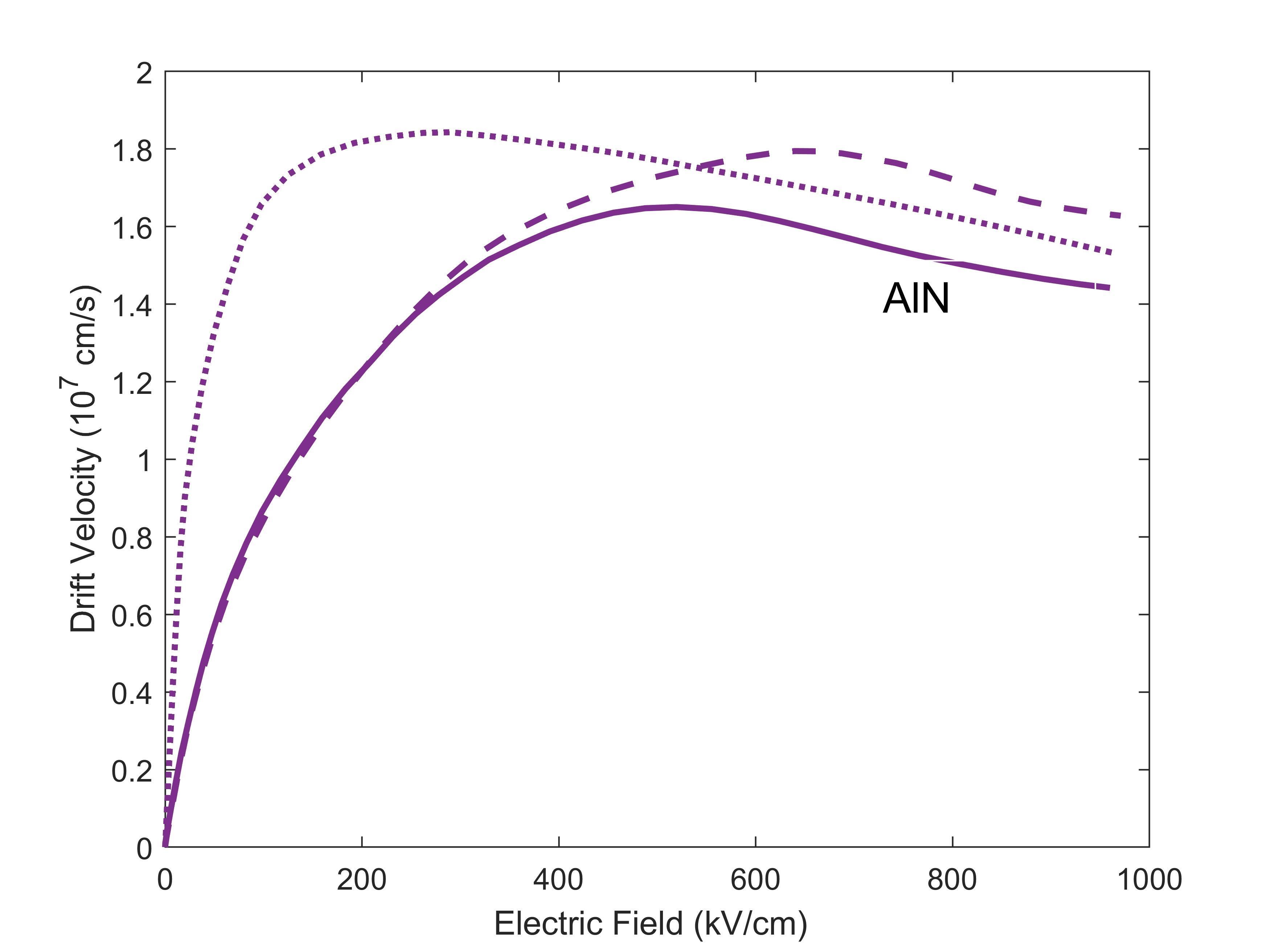}
    \caption{}
    \label{fig: first}
\end{subfigure}
\hfill
\begin{subfigure}{0.49\textwidth}
    \centering
    \includegraphics[width=\textwidth]{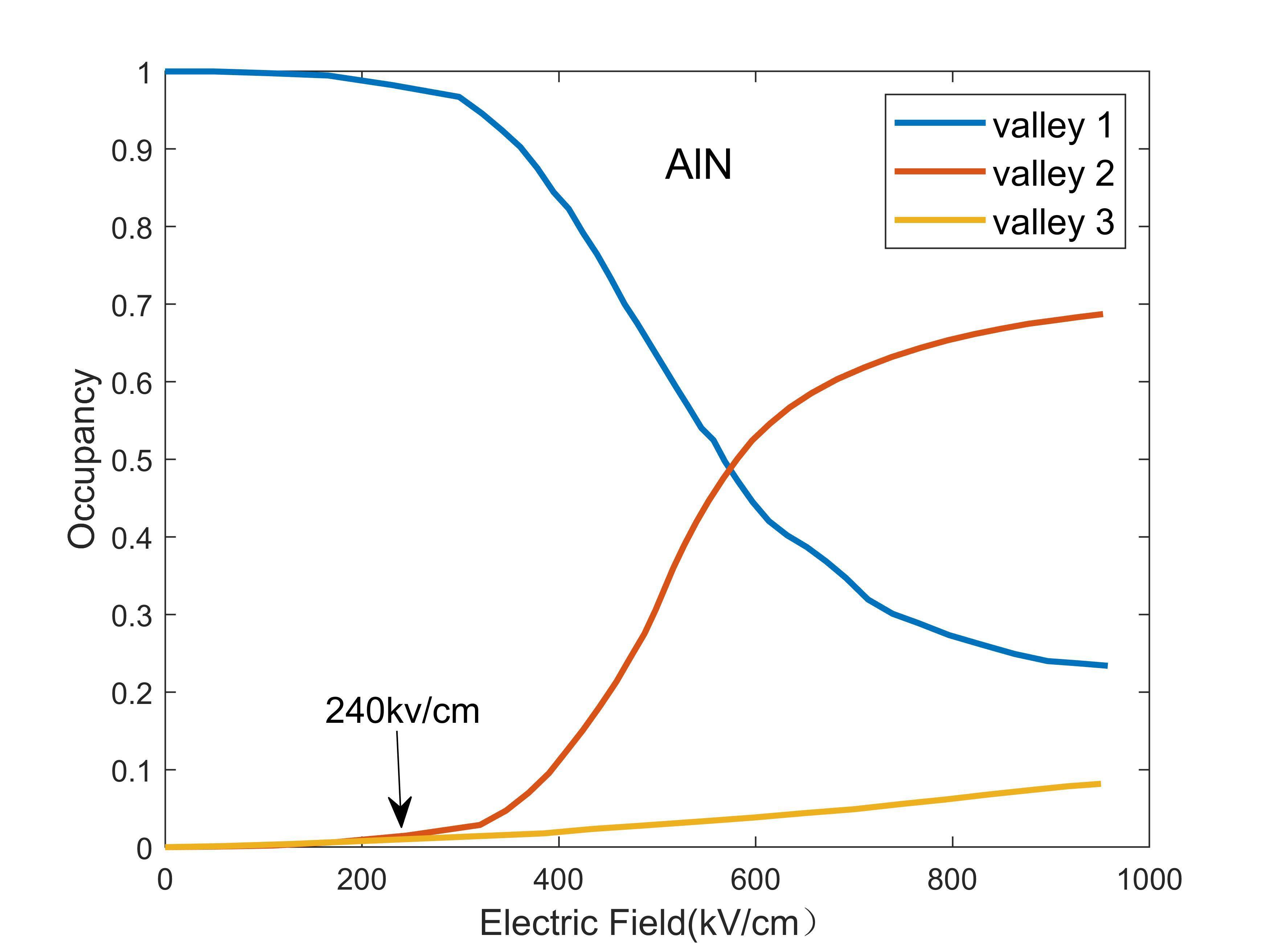}
    \caption{}
    \label{fig: second}
\end{subfigure}
\hfill
\begin{subfigure}{0.49\textwidth}
    \centering
    \includegraphics[width=\textwidth]{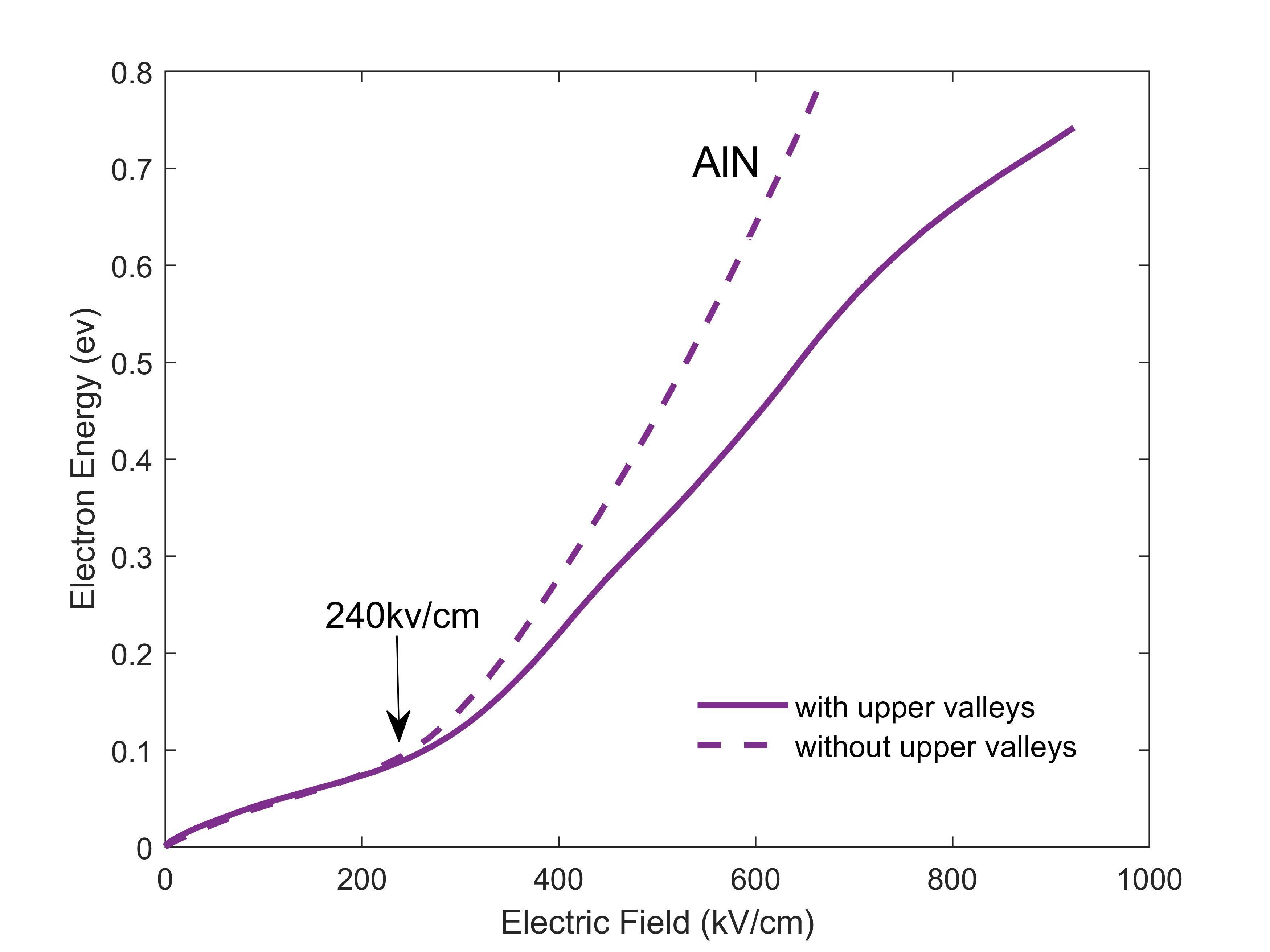}
    \caption{}
    \label{fig: second}
\end{subfigure}
\caption{(a) shows the velocity as a function of the electric field associated with bulk wurtzite AlN with n=$10^{17}cm^{-3}$ at 300K. The solid line is all considered in the 3-valley model, the dashed line is without upper valleys, and the dotted line is without longitudinal polar optical phonon scattering in the central valley. (b) shows the valley occupancy as a function of the electric field associated with bulk wurtzite AlN. (c) shows the average electron energy as a function of the electric field with and without upper valleys associated with bulk wurtzite AlN.} 
\label{fig: figures}
\end{figure}

Since the material properties control the magnitude and position of the velocity overshoot it should be possible to tune these by alloying the binary materials. The properties of the ternary nitrides are expected to lie between the binaries. Therefore, the ternary alloys of Al$_x$Ga$_{1-x}$N are investigated as a function of alloy composition as follows. All of the parameters listed in Table 1 are varied linearly with composition except for the energy position of the bands which was discussed in section 2. Fig.6a shows that the LPOP emission scattering rate increases with increasing Al composition. However, the rate of change of the scattering rate lessens with increasing Al, which is consistent with the trend of the rate of change of the peak velocity amplitude when IV scattering is not considered.

This shows that LPOP scattering is an important mechanism for electron velocity characteristics. Including IV scattering for the ternary alloy is affected by the changing position of the upper valleys with Al composition as seen in Fig.2 which is reflected in more complex behavior. The variation of IV scattering rate with Al composition is shown in Fig.6b, decreasing at the beginning and then increasing. Although the turning point occurs at x=0.38, there is an obvious gap between x=0.5 and x=0.625.

\begin{figure}
\centering
\begin{subfigure}{0.49\textwidth}
    \centering
    \includegraphics[width=\textwidth]{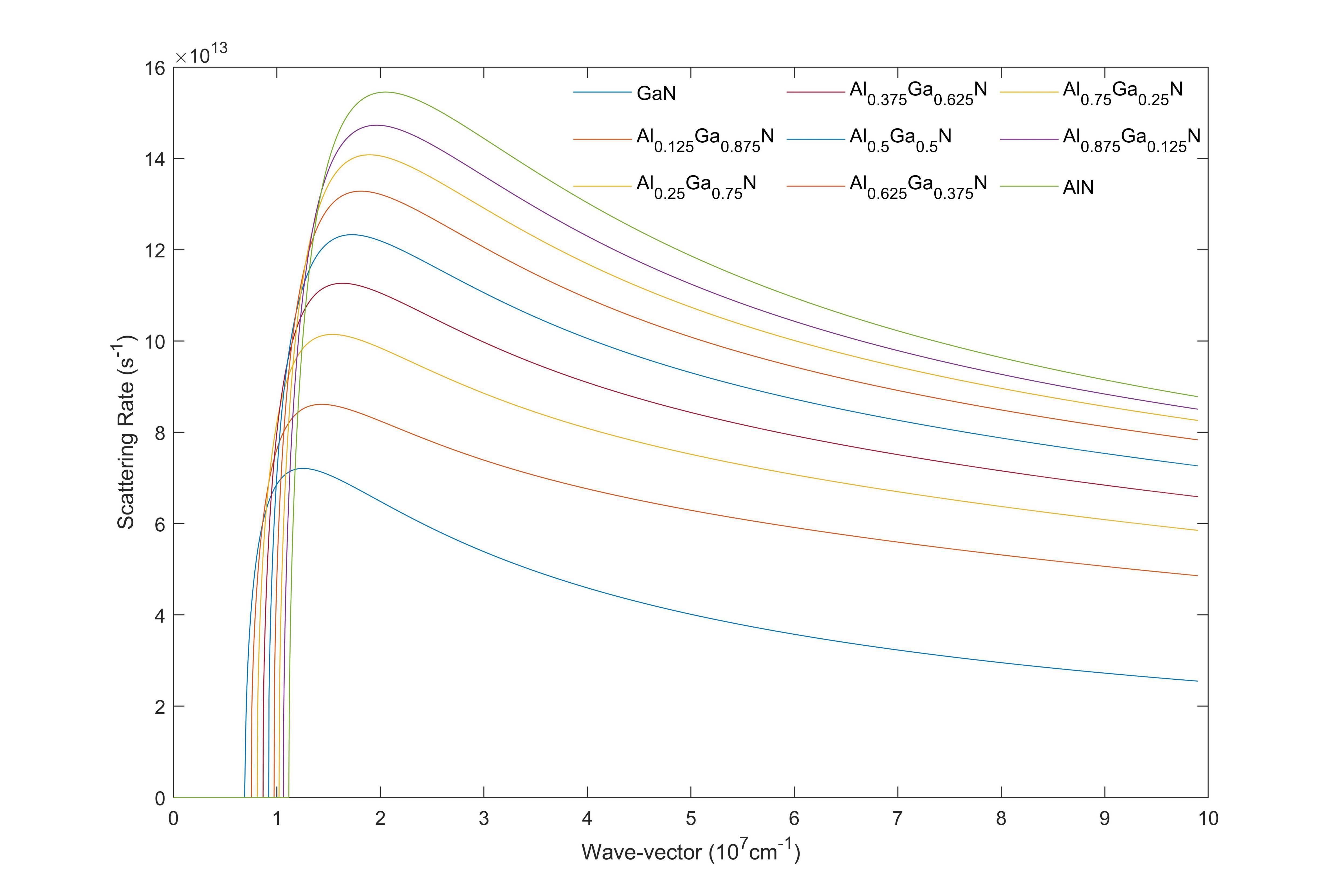}
    \caption{}
    \label{fig: first}
\end{subfigure}
\hfill
\begin{subfigure}{0.49\textwidth}
    \centering
    \includegraphics[width=\textwidth]{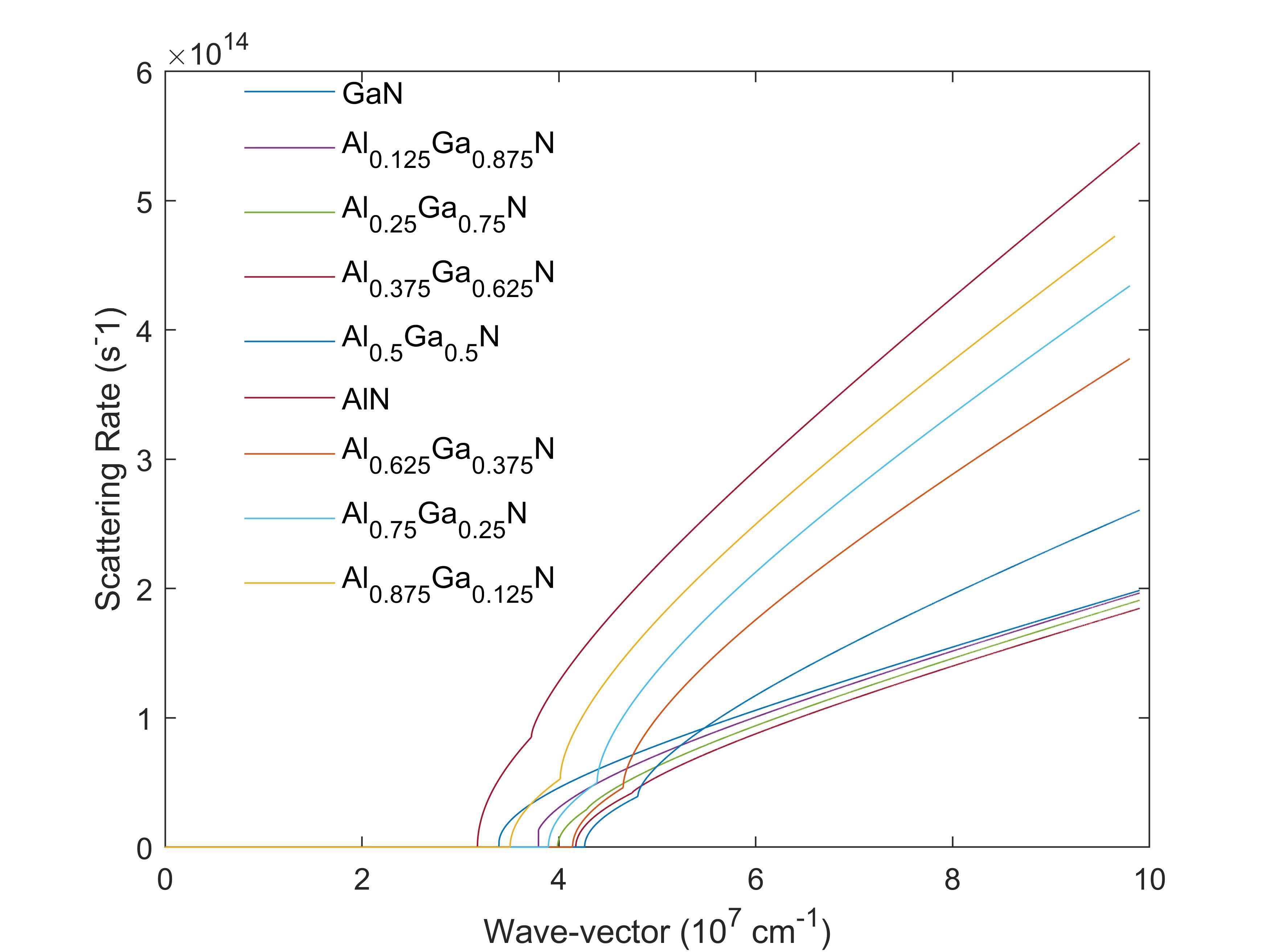}
    \caption{}
    \label{fig: second}
\end{subfigure}
\caption{(a) shows the longitudinal polar optical phonon emission scattering rate of Al$_x$Ga$_{1-x}$N. When the Al composition becomes larger, the spacing between the scattering rate curves gradually becomes smaller, which means that the rate of change of the scattering rate becomes smaller. (b) shows the sum of the inter-valley emission scattering rate between the center valley and upper valleys. When x=0.125$\sim$0.375, the scattering rate slightly decreases and then begins to increase. The most significant increase is between x=0.5 and x=0.625.} 
\label{fig: figures}
\end{figure}

Fig.7a shows the electron velocity as a function of the F field of Al$_x$Ga$_{1-x}$N for different Al compositions. Fig.7b shows the velocity-field curves neglecting IV scattering showing that the IV scattering is not dominant. The electron velocity as a function of the field exhibits a reasonably smooth change from the GaN and AlN binaries however the change of the characteristics is nonlinear with the composition. We also notice that when IV scattering is considered, the rate of change of the peak velocity has an unusual increase rather than a continuous decrease(compared to others) for Al compositions between  0.5 and 0.625. However, this is not the case when IV scattering is omitted. In the 3-valley model for ternary alloys, once the positions of the upper valleys change, the characteristics may have unconventional changes.

\begin{figure}
\centering
\begin{subfigure}{0.49\textwidth}
    \centering
    \includegraphics[width=\textwidth]{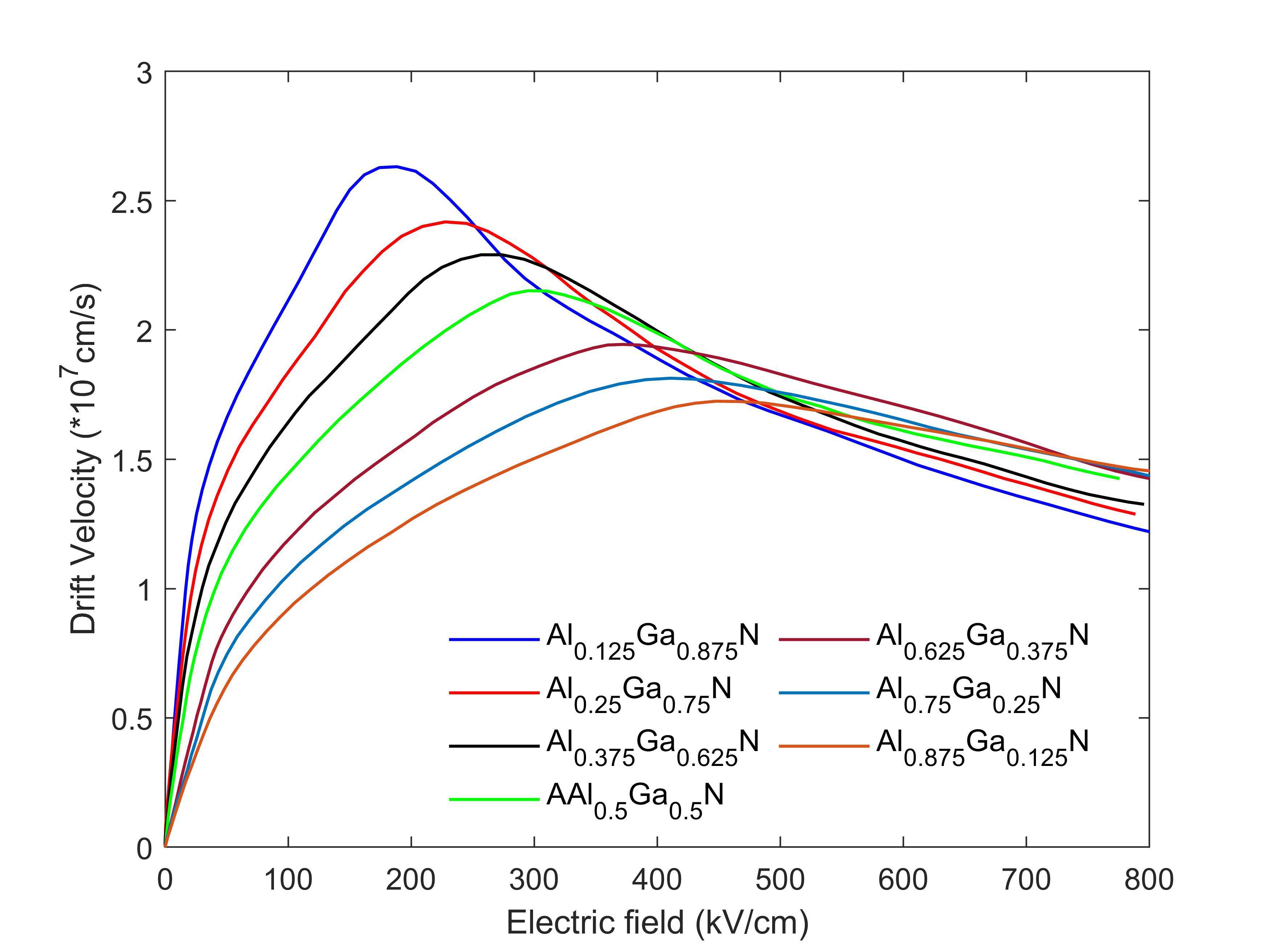}
    \caption{}
    \label{fig: first}
\end{subfigure}
\hfill
\begin{subfigure}{0.49\textwidth}
    \centering
    \includegraphics[width=\textwidth]{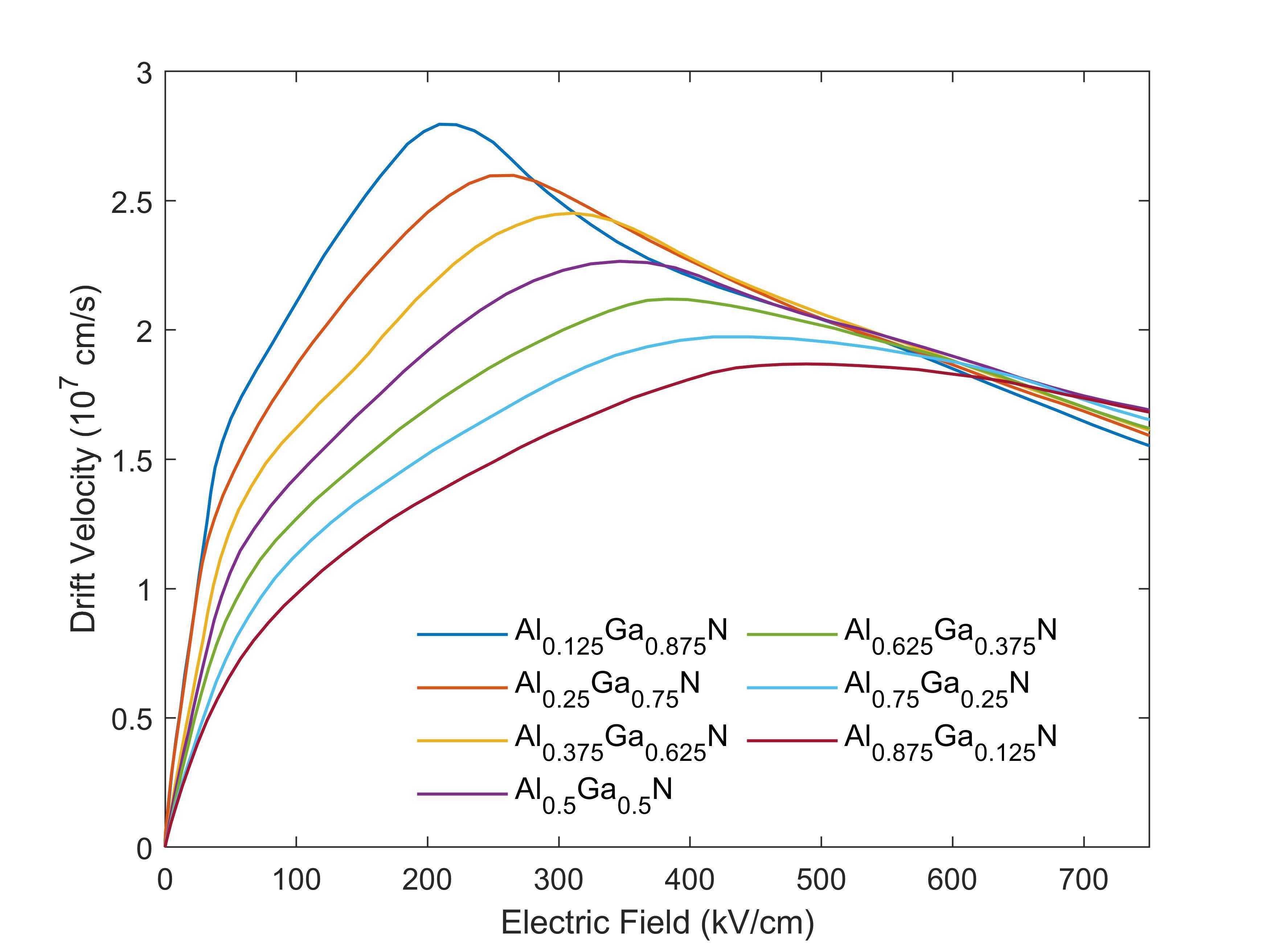}
    \caption{}
    \label{fig: second}
\end{subfigure}
\caption{(a) shows the electron velocity as a function of the electric field of bulk Al$_x$Ga$_{1-x}$N for various x. (b) shows the velocity curves omitting inter-valley scattering for various x. Doping concentration=$10^{17}cm^{-3}$, T=300K.} 
\label{fig: figures}
\end{figure}

\newpage

\section{Modeling of superlattices}

\subsection{Band structure}
GaN/AlGaN SLs exhibit a unique behavior where the energy levels of individual wells merge due to the periodic potential created by alternating layers. As a result, minibands are formed in these structures. Using the lattice constants of GaN and AlN, we design and present the SLs in terms of monolayer(ML) thicknesses of 0.26 nm and 0.25 nm, respectively[28]. The monolayer thicknesses of Al$_x$Ga$_{1-x}$N as a function of x are in the range of 0.25nm$\sim$0.26nm. The SLs discussed in this paper consider the quantum well(QW) width($L_{QW}$) to be equal to the quantum barrier(QB) width($L_ {QB}$).

We used the Schrödinger wave equation to solve for the periodic potential formed by the alternating layers of GaN and AlGaN in the growth direction (z-axis). The eigenfunctions $\psi$ are constructed from a sequence of plane waves $k_i$ and remain continuous at the material interfaces and exhibit smoothness at these interfaces, ensuring current conservation as $\frac{1}{m^*}\frac{ d\Psi}{dz}$[29]. The wavefunction reflects the lattice periodicity L, such that $\Psi(z+L)=\Psi(z)exp(ikL)$. The boundary conditions are periodic, applicable to infinite superlattice, suitable for more than 10 periods.
The period of superlattice L is represented as L = $L_{QW}$ + $L_{QB}$. 

The superlattice band structure depends on the energy offset between the conduction band in the QW and QB material and the band structure of the conduction band of each material. In this calculation, we take a 1-band model with a non-parabolic band structure and match the traveling waves as described above. A more complex band structure including higher additional bands would involve matching traveling waves for these as well to form higher bands existing through the superlattice. The band structure of the lowest conduction band in each material coalesced into a miniband through the superlattice which has a band structure determined by the individual constituent materials and their thicknesses. The resulting miniband has an energy position, energy width and a band structure that is determined by the material constituents, in a 1-band non-parabolic approximation, and the thickness of their layers. In the wurzite materials,  piezo-F fields and spontaneous polarisation contribute an in-built F field within the structure and charged donor ions and  free space charge in the QW contribute to charges that distort band edges and are included in the Poisson Solver. The simulation incorporates these effects by linking the Schrödinger wave equation with a Poisson solver and solves them self-consistently through iterative solutions. Nextnano is employed to simulate superlattice band structure using a 1-band conduction band model for each material comprising the superlattice and integrates a non-linear Poisson equation solver that employs a preconditioned conjugate gradient (PCG) iterative technique for calculations[30]. The Poisson-Schrödinger equation undergoes a self-consistent resolution under periodic boundary conditions, necessitating a repetitive resolution procedure. Electrostatic potentials and wave patterns are recurrently revised until they achieve self-consistency, facilitating precise estimations of band extremities and other attributes. The parameters align with those of the bulk materials previously. Using a 1-band model for each material comprising the superlattice. The effective masses of electrons we use here are in the z-direction.

We calculated the conduction band edges of n-doped GaN/Al$_x$Ga$_{1-x}$N SLs with different x and widths. The calculation considers the band shift due to strain, piezoelectric, and spontaneous polarisation effects.
We consider GaN/AlGaN superlattices with the AlGaN barrier under tensile strain and the GaN well lattice-matched to the substrate (sapphire). The parameters used for these calculations were obtained from references[31] and [32]. The conduction band offset of GaN/AlN material is 1.2eV and varies linearly with x[33]. The Si donor energy in GaN is 30meV[34], and the Si donor energy in Al$_x$Ga$_{1-x}$N is taken from 30mev to 250mev in a very non-linear manner[34][35]. We show the conduction band edges, Fermi level, donor energy level and the first miniband energy position and width of GaN/Al$_{0.2}$Ga$_{0.8}$N 4MLs/4MLs and 6MLs/6MLs in Fig.8. The QB heights for these two cases without the in-build electric field effects are 0.47eV and including in-build electric field go to 0.29eV. The miniband energy position changes because of the change of L, the confinement upshift at the center of the QW is 0.13eV(4MLs/4MLs) and 0.12eV(6MLs/6MLs). The dimensions of $L_{QW}$ and $L_{QB}$ play a major role in determining the miniband width. In the case of 4MLs/4MLs, the miniband width is 310meV, whereas the SL of 6MLs/6MLs is just 112meV in width. The Al content in the QB also impacts the miniband properties. The bandwidth of the miniband for GaN/Al$_{0.5}$Ga$_{0.5}$N 4MLs/4MLs is 180meV while for GaN/AlN 4MLs/4MLs, it is 65meV, showing that increasing x narrows the energy bandwidth. Since we are interested in high field, high velocity transport we need a significant energy width of the miniband so we focus the study on low x content and short periods.

\begin{figure}
\centering
\begin{subfigure}{0.49\textwidth}
    \includegraphics[width=\textwidth]{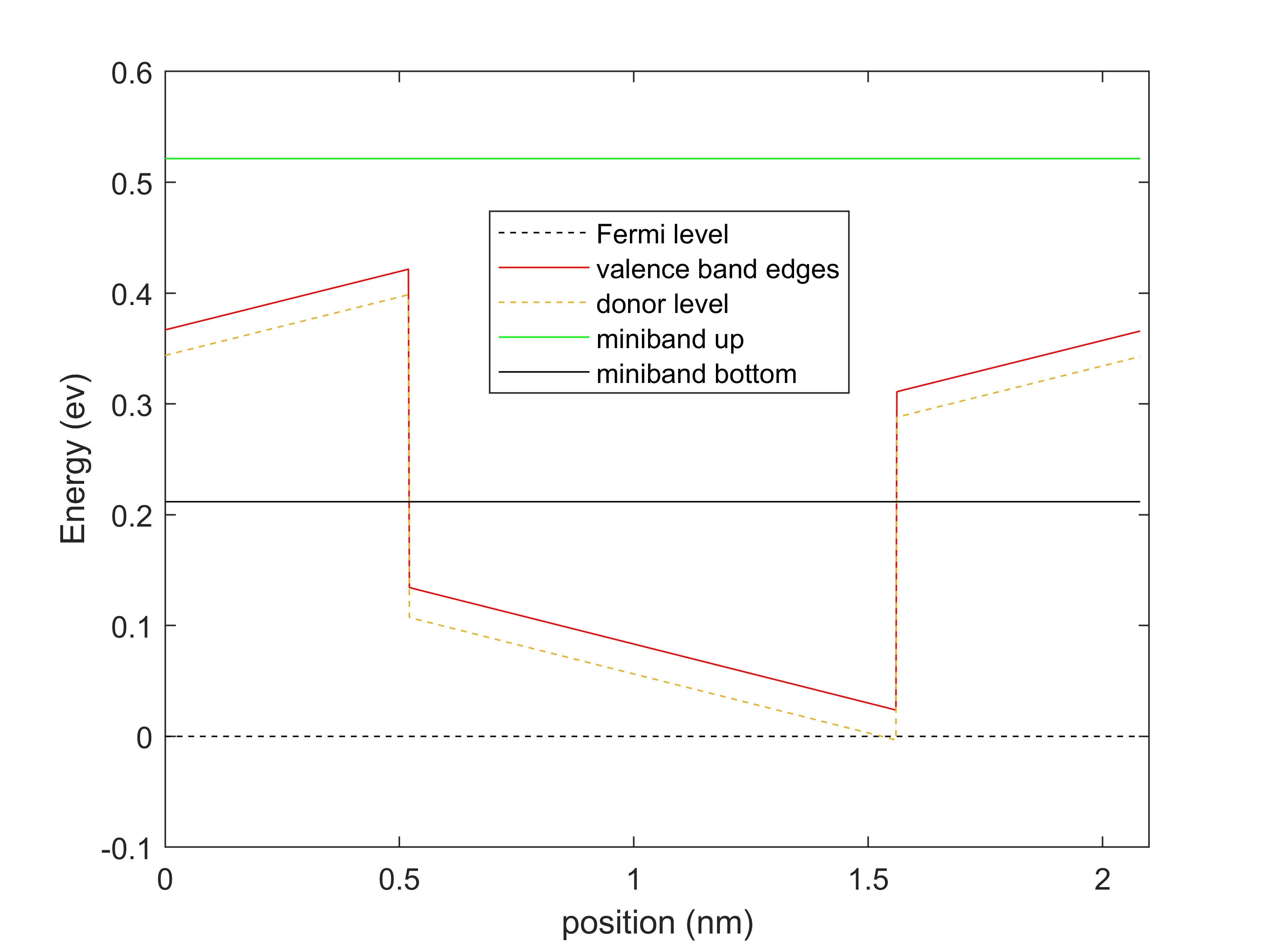}
    \caption{}
    \label{fig: first}
\end{subfigure}
\hfill
\begin{subfigure}{0.49\textwidth}
    \includegraphics[width=\textwidth]{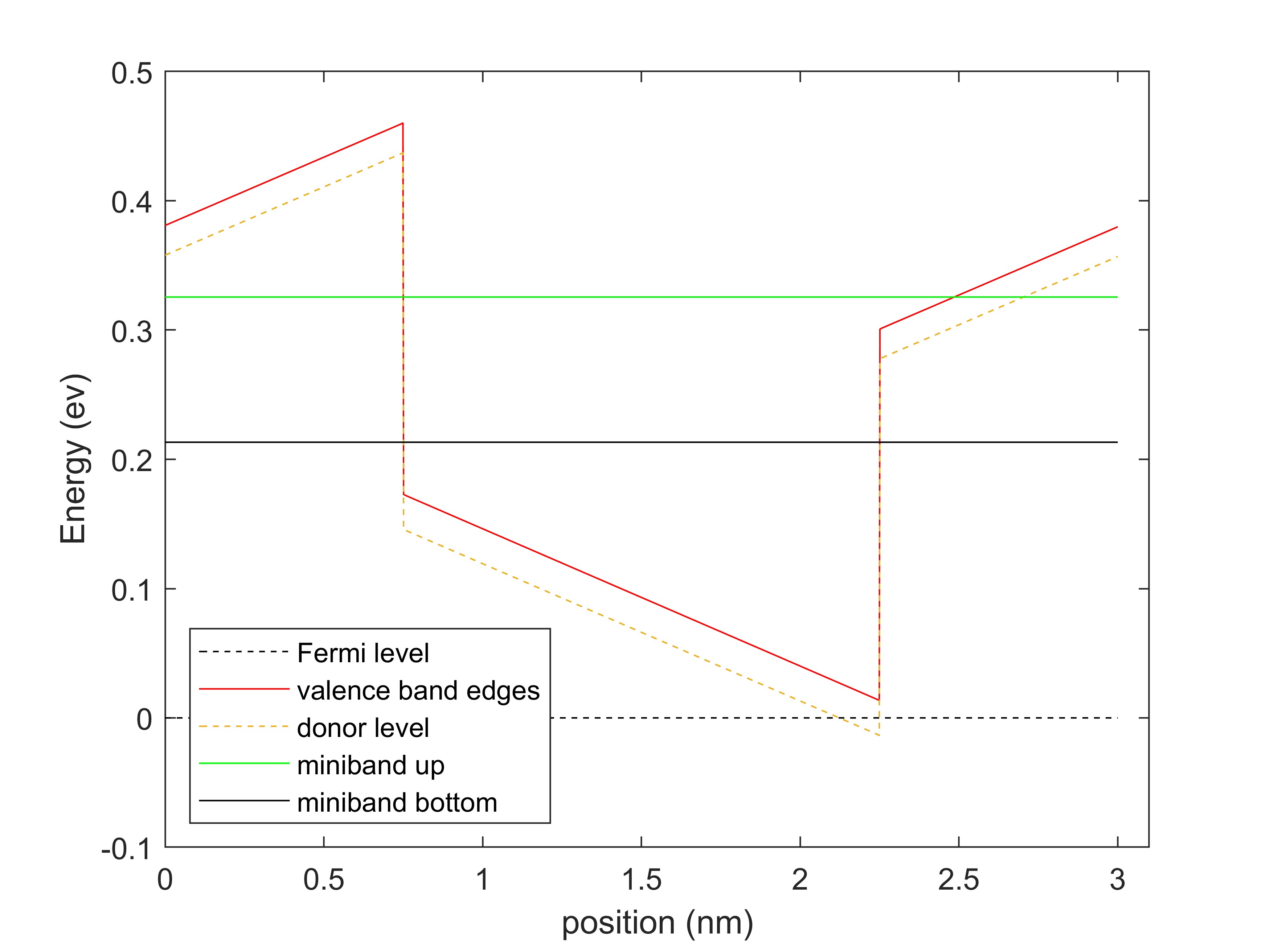}
    \caption{}
    \label{fig: second}
\end{subfigure}
\caption{The conduction band edges, donor energy levels, Fermi levels and miniband energy levels of GaN/Al$_{0.2}$Ga$_{0.8}$N 4MLs/4MLs(a) and 6MLs/6MLs(b) with N$_A$ = $10^{17}cm^{-3}$, T=300K} 
\label{fig: figures}
\end{figure}

Fig.9 illustrates the dispersion of the hole miniband across various  $L_{QW}$=$L_{QB}$. As L becomes larger, the dispersion of the miniband flattens, indicative of states being localized within the QWs. Conversely, as L decreases, the dispersion of the miniband increases, signifying a greater extent of delocalization of the miniband states. The focus of this study is on short-period superlattices where L is small and the effective mass of the miniband is small which should improve electron transport in the vertical direction. The effective mass of the miniband can be extracted from the dispersion curve. Close to the $\Gamma$ point, a parabolic effective mass can be used while for large k values of the miniband, non-parabolicity can be incorporated to fit the dispersion more accurately. The parabolic effective mass of some representative short-period superlattice minibands is shown in Table 2. For small L, the effective mass approaches values intermediate between the electron masses of the QW and QB while for large L, the effective mass increases considerably, even surpassing that of QB. Furthermore, increasing the Al content also leads to a significant increase in this value reflecting the higher effective mass of Al$_x$Ga$_{1-x}$N in z-direction with higher x.

\begin{figure}[!t]
\graphicspath{ {./figure_300/} }
\centering{\includegraphics[width=0.5\textwidth]{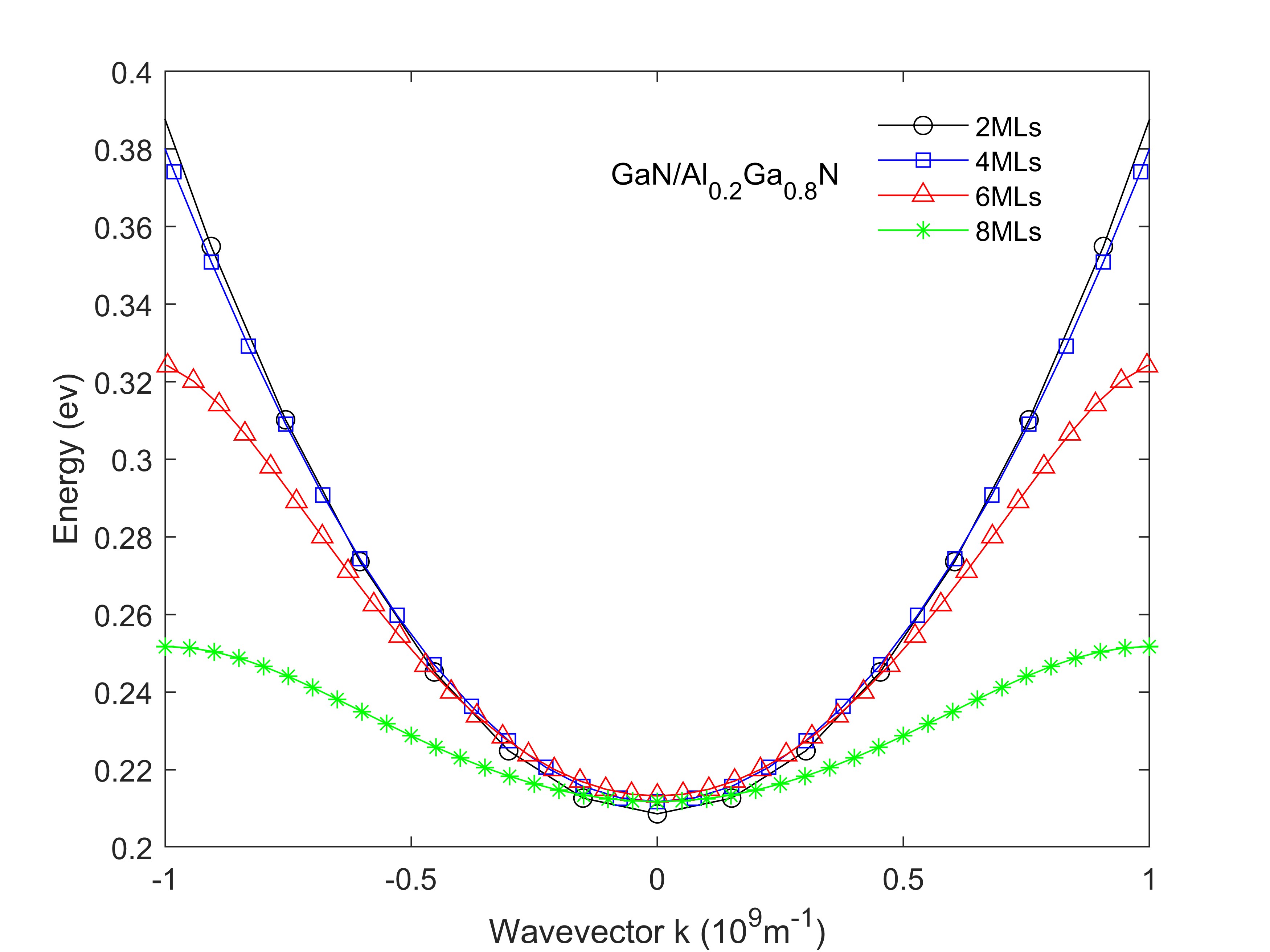}}
\caption{ The electron miniband dispersion of GaN/Al$_{0.2}$Ga$_{0.8}$N SLs with different period L.}
\end{figure}

\begin{table}
 \caption{The effective mass of GaN/Al$_x$Ga$_{1-x}$N with different thicknesses.}
 \centering
  \begin{tabular}[htbp]{@{}lllll@{}}
    \hline
    Material & 2MLs & 4MLs & 6MLs & 8MLs \\
    \hline
    GaN/AlN     &0.3$m_0$   &0.78$m_0$  &4.75$m_0$  &43.6$m_0$ \\
    \hline
    GaN/Al$_{0.5}$Ga$_{0.5}$N     &0.24$m_0$   &0.31$m_0$  &0.63$m_0$           &2.06$m_0$ \\
    \hline
    GaN/Al$_{0.2}$Ga$_{0.8}$N     &0.21$m_0$   &0.22$m_0$  &0.24$m_0$           &0.36$m_0$ \\
    \hline
  \end{tabular}
\end{table}

In bulk materials, the non-parabolic factor can be more pronounced in specific regions of the band structure, such as at high miniband energies or when the band gap is small. In contrast, SL structures, with their periodic arrangement of alternating materials, may exhibit different non-parabolic behavior compared to bulk materials. The mini-Brillouin zones formed within SLs and the confinement effects can influence the electronic band structure, leading to unique non-parabolic characteristics. By precisely determining the range of applicability of the parabolic approximation, we can effectively model and analyze electron behavior within the miniband, enabling a more meaningful simulation of electron transport in the system.

Fig.10 illustrates a comparison of E-k dispersion curves for several GaN//Al$_x$Ga$_{1-x}$N superlattice minibands compared with parabolic approximations.

Fig.10(a),(b) and (c) presents the non-parabolic behavior under different Al contents, while Fig.10(b), (d), and (e) show the non-parabolic approximations for different L of the SLs. Notably, the E-k curves demonstrate a clear trend of deviating from the parabolic approximation as the Al content and superlattice L increase. Consequently, unlike in bulk materials where the region up to k=0.5$\times10^{9}m^{-1}$  is commonly used in electron transport calculations, in the case of minibands in SLs, it is more appropriate to redefine the calculation region for up to k=0.4$\times10^{9}m^{-1}$.

\begin{figure}
\centering
\begin{subfigure}{0.32\textwidth}
    \includegraphics[width=\textwidth]{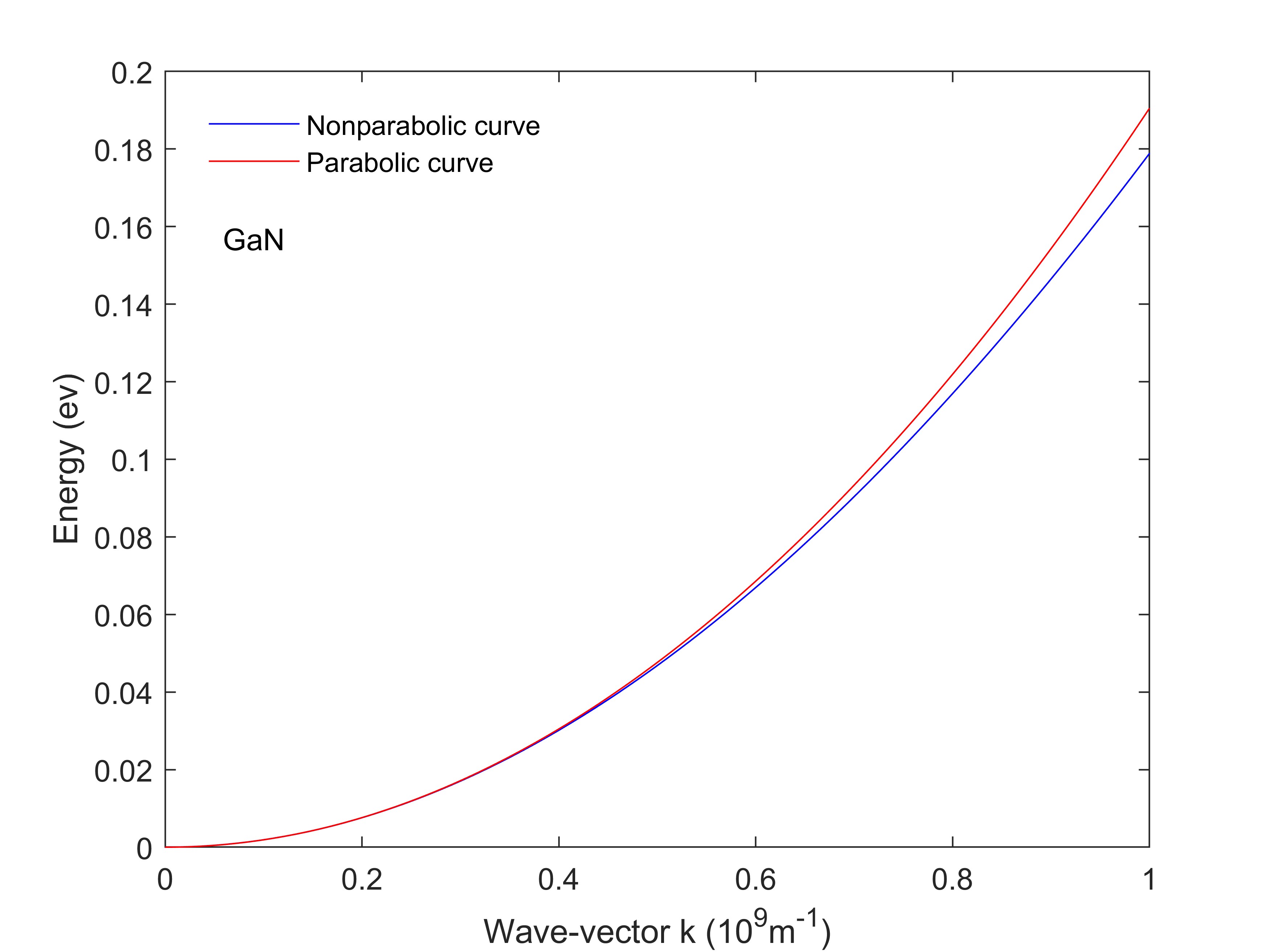}
    \caption{}
    \label{fig:first}
\end{subfigure}
\hfill
\begin{subfigure}{0.32\textwidth}
    \includegraphics[width=\textwidth]{figure_300/Fig.11a.jpg}
    \caption{}
    \label{fig:second}
\end{subfigure}
\hfill
\begin{subfigure}{0.32\textwidth}
    \includegraphics[width=\textwidth]{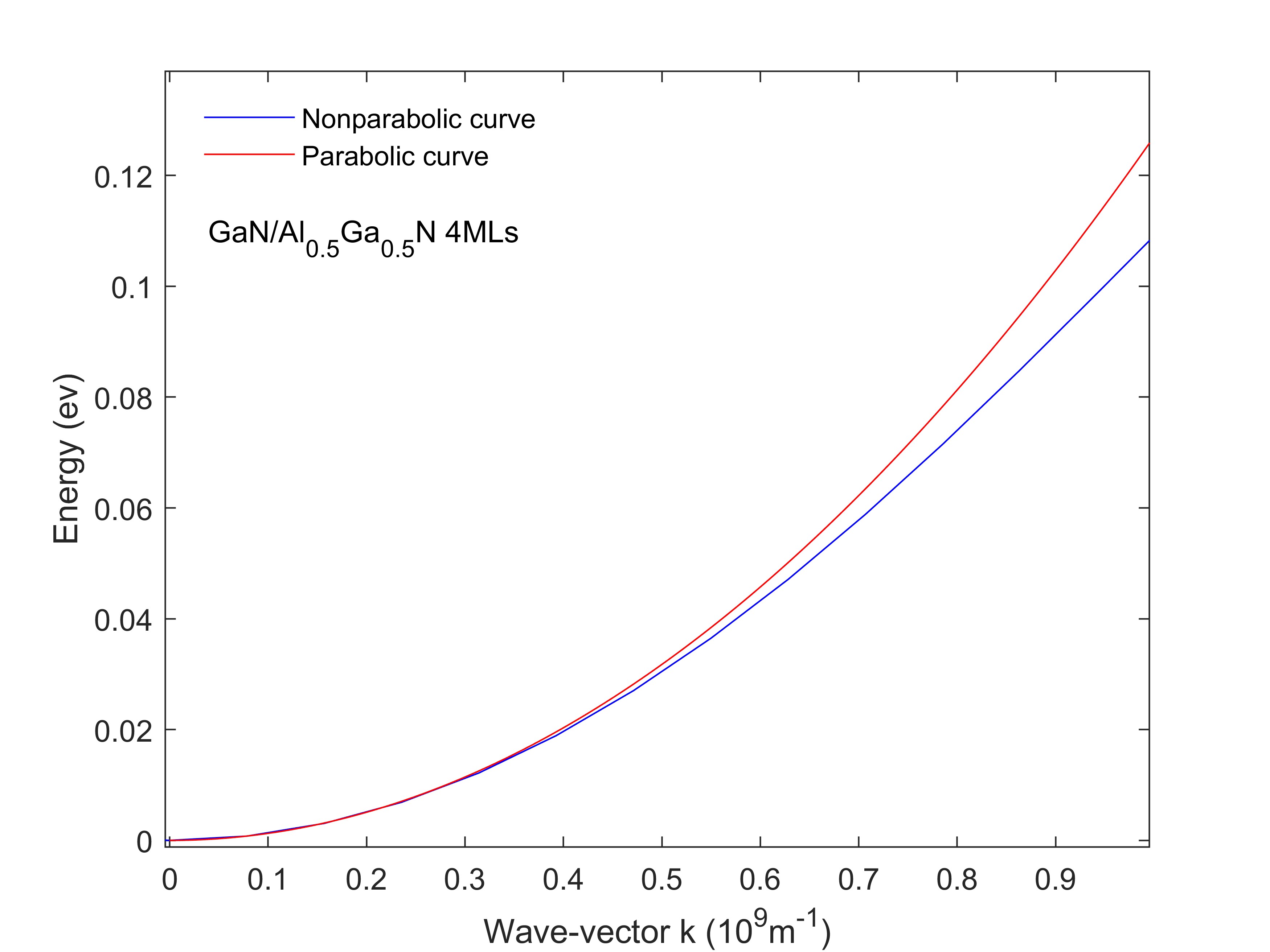}
    \caption{}
    \label{fig:second}   
\end{subfigure}
\centering
\begin{subfigure}{0.32\textwidth}
    \includegraphics[width=\textwidth]{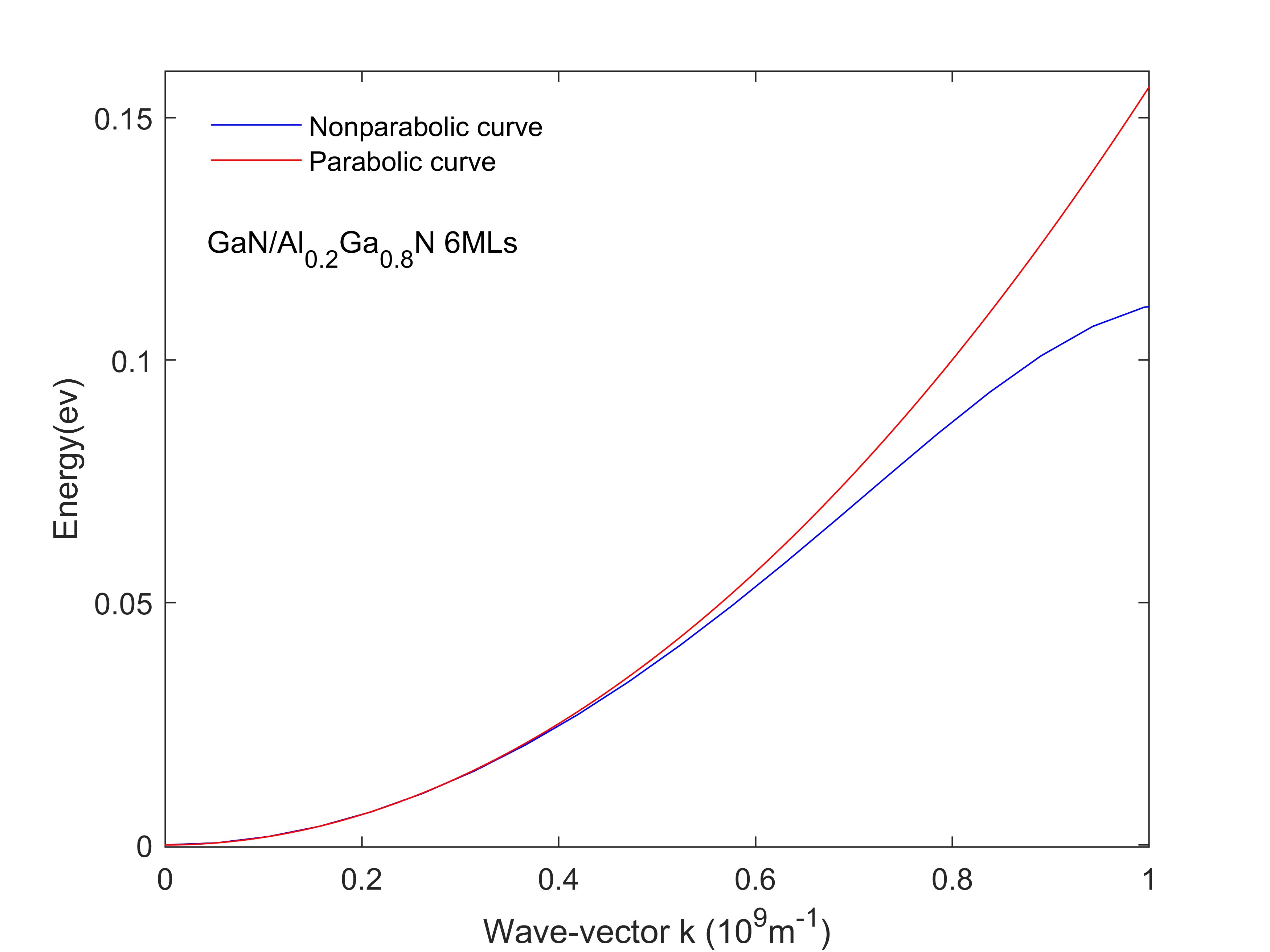}
    \caption{}
    \label{fig:second}   
\end{subfigure}
\hfill
\begin{subfigure}{0.32\textwidth}
    \includegraphics[width=\textwidth]{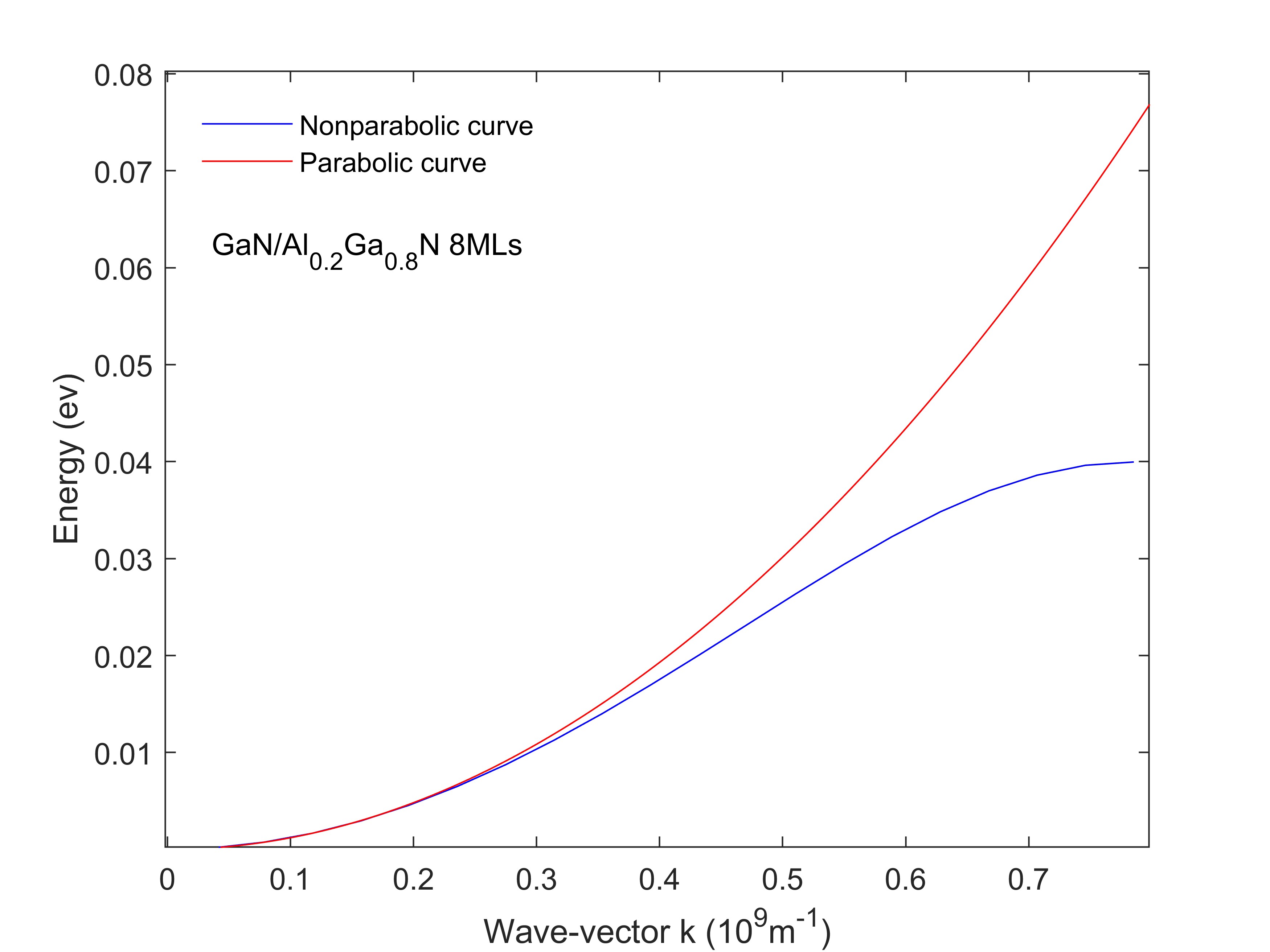}
    \caption{}
    \label{fig:second}   
\end{subfigure}

\caption{Comparison of the parabolic and non-parabolic cases. (a) shows the energy–wavevector (E–k) dispersion curves of GaN.  (b) shows the curves of 4MLs GaN/Al$_{0.2}$Ga$_{0.8}$N. (c) shows the curves of 4MLs GaN/Al$_{0.5}$Ga$_{0.5}$N. (d) shows the curves of 6MLs GaN/Al$_{0.2}$Ga$_{0.8}$N. (e) shows the curves of 8MLs GaN/Al$_{0.2}$Ga$_{0.8}$N} 
\label{fig:figures}
\end{figure}

\subsection{Monte Carlo results of superlattices}

The electron transport in SLs is influenced by several factors: the strength of the F field, the effective mass of superlattice miniband in the growth direction, non-parabolicity factors of the miniband dispersions and various scattering mechanisms. In this study, the focus is on intraband scattering for electrons traveling in the lowest miniband, neglecting weak interband scattering to higher minibands[36]. Based on the analysis of bulk material presented in Section II we recognize the significant role of LPOP scattering. For GaN/Al$_x$Ga$_{1-x}$N SLs particularly for low x, it is appropriate to consider that the phonons involved will be bulk GaN phonons (91.2meV) that will be continuous throughout the SL. Therefore we evaluate the LPOP scattering rate in SLs by considering the electron effective masses of the miniband within the SLs and its dispersion (see Fig.11) combined with bulk GaN phonons. In Fig.11a, we compare the LO phonon scattering rate of SLs with varying Al content to that of bulk GaN. Fig.11b shows the comparison of the SLs scattering rate for L. Clearly, the LPOP scattering rate in SLs increases with increasing Al composition and increasing L.

\newpage

\begin{figure}
\centering
\begin{subfigure}{0.49\textwidth}
    \includegraphics[width=\textwidth]{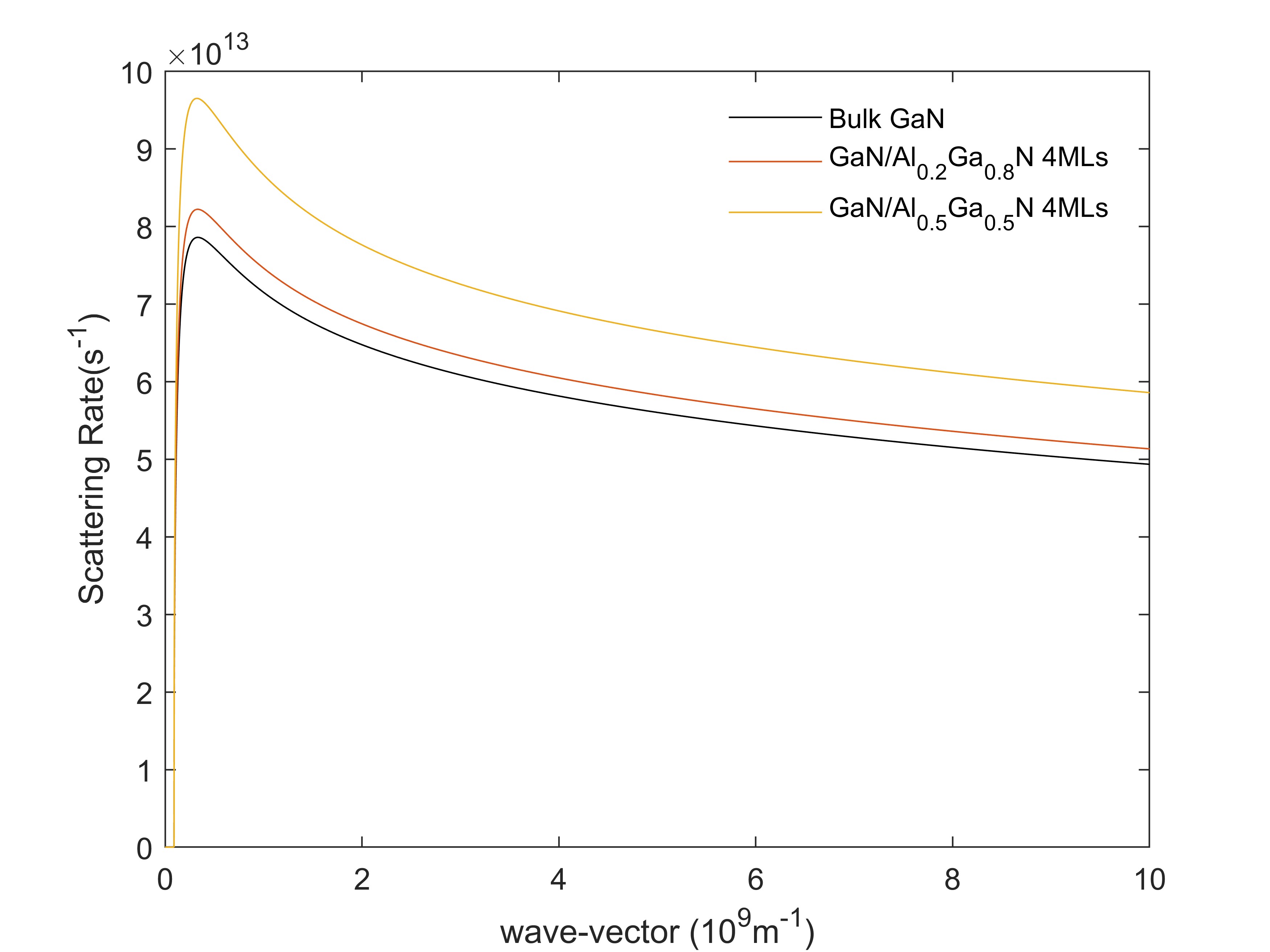}
    \caption{}
    \label{fig:first}
\end{subfigure}
\hfill
\begin{subfigure}{0.49\textwidth}
    \includegraphics[width=\textwidth]{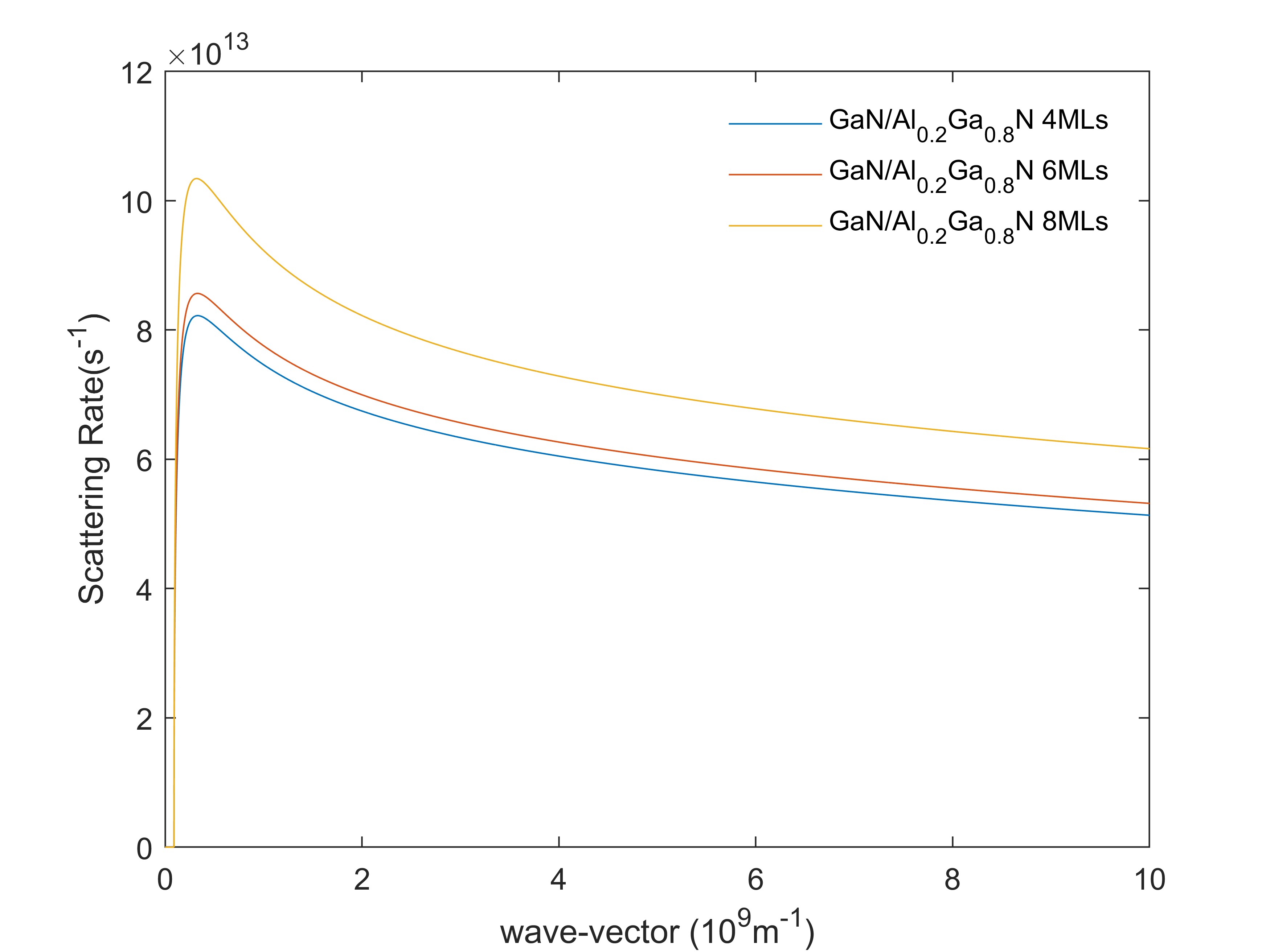}
    \caption{}
    \label{fig: second}
\end{subfigure}
\caption{(a) shows the electron velocity as a function of the electric field of bulk Al$_x$Ga$_{1-x}$N for various x. (b) shows the velocity curves omitting inter-valley scattering for various x. Doping concentration=$10^{17}cm^{-3}$, T=300K.} 
\label{fig: figures}
\end{figure}

Additional intraband scattering processes should also be considered for the n-doped GaN/AlGaN SLs. Ionized impurity scattering is believed to have an important detrimental role in p-dope GaN/AlGaN SLs where the acceptors are very deep[37]. Since the donors in GaN and Al$_x$Ga$_{1-x}$N with low x, have a low binding energy, comparable to the thermal energy they are expected to be ionized throughout the superlattice as they would be in the bulk material so the ionized impurity scattering should not be very different from the bulk material. Piezoelectric scattering in the SLs is taken as the bulk value which is small in magnitude. Previous calculations[15] have demonstrated that acoustic scattering for transport in GaAs-based SLs is not significantly altered compared to bulk materials. Consequently, we can make the assumption that in GaN materials, the phonon scattering remains comparable in both SLs and bulk, except for the effect of the electron effective masses and miniband dispersion effects. The first miniband for L from 4MLs to 8MLs provides a wide energy width with high energy states and will have scattering similarly to the bulk materials. It must be remembered that we neglect IV scattering but below this threshold(which was studied in section II)  similar scattering as in the bulk material will occur with LPOP scattering being dominant. Additionally, the considerable energy gap between the first miniband and higher-level minibands reduces inter-band scattering, which is reduced in magnitude compared to intravalley scattering anyway[36]. Thus, we use a Monte Carlo transport analysis within the 1-band model to calculate electron transport within the GaN/Al$_x$Ga$_{1-x}$N SLs.  We present the velocity as a function of the F field for GaN/Al$_x$\allowbreak Ga$_{1-x}$N SLs under various x compositions (Fig.12a) and different L (Fig.12b) in a few specific cases in Fig.12. Compared to the bulk material GaN (in this case, peak velocity=3.2$\times10^{7}$cm/s at F=234kV/cm), GaN/Al$_{0.2}$\allowbreak Ga$_{0.8}$N 4MLs has the peak velocity of 2.95$\times10^{7}$cm/s at 248kV/cm, while the peak velocity of GaN/Al$_{0.5}$\allowbreak Ga$_{0.5}$N 4MLs is 2.6$\times10^{7}$cm/s, obtained at 357kV/cm. When we increase the L of GaN/Al$_{0.2}$\allowbreak Ga$_{0.8}$N from 4MLs to 8MLs, the peak velocity further drops to 2.4$\times10^{7}$cm/s and occurs at 430kV/cm. It can be observed that increasing the content of the Al component or increasing the L of SLs has a large influence on the drift velocity. The velocity overshoot of the SLs becomes less pronounced for high Al content in the QB or for large L.

\begin{figure}
\centering
\begin{subfigure}{0.49\textwidth}
    \includegraphics[width=\textwidth]{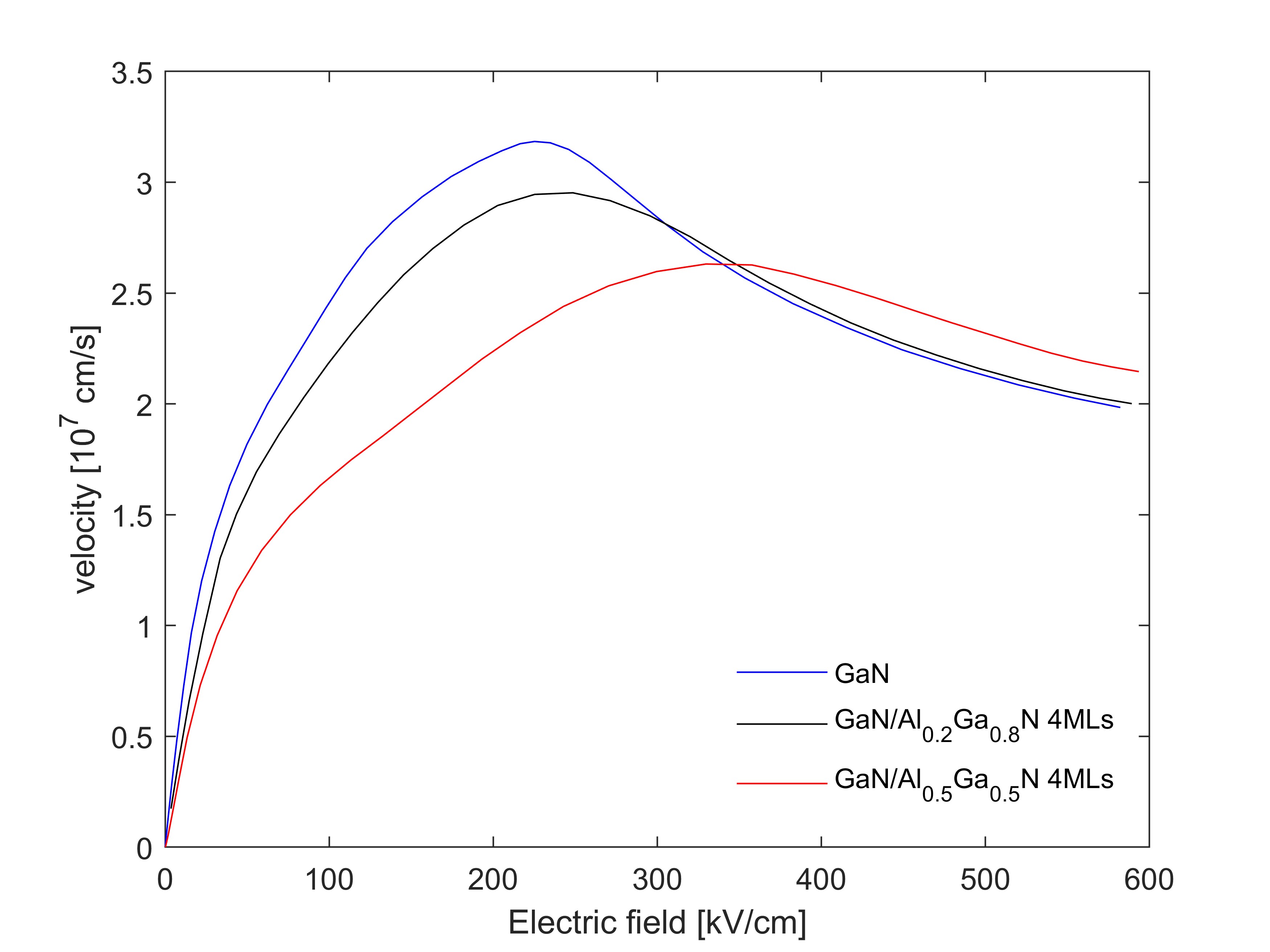}
    \caption{}
    \label{fig: first}
\end{subfigure}
\hfill
\begin{subfigure}{0.49\textwidth}
    \includegraphics[width=\textwidth]{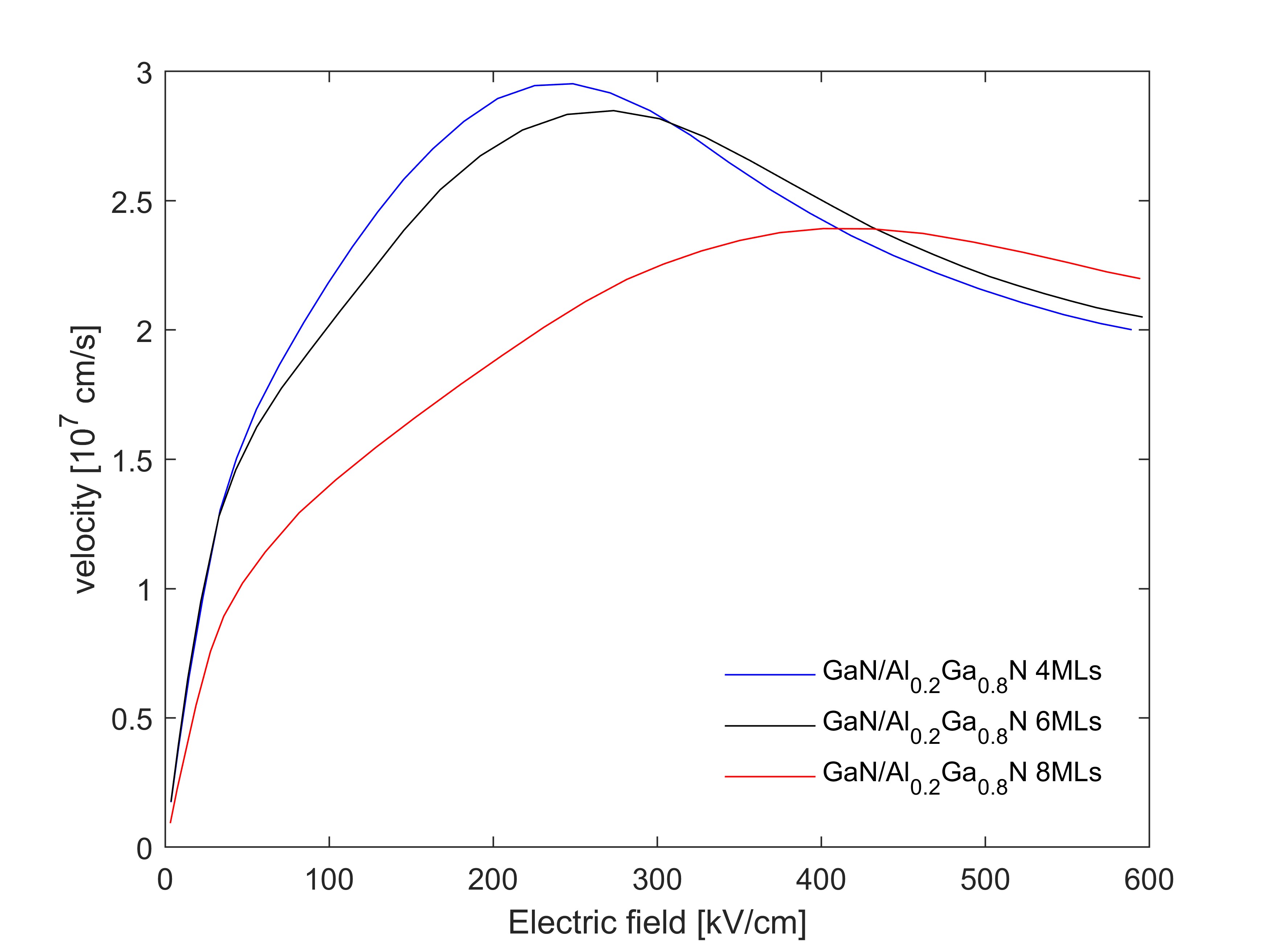}
    \caption{}
    \label{fig: second}
\end{subfigure}
\caption{(a) shows the electron velocity as a function of the electric field of  GaN/Al$_x$Ga$_{1-x}$N for various x(0.2 and 0.5). (b) shows the velocity curves as a function of the electric field of  GaN/Al$_{0.2}$Ga$_{0.8}$N SLs for various thicknesses (4MLs, 6MLs and 8MLs). Doping concentration=$10^{17}cm^{-3}$, T=300K.} 
\label{fig: figures}
\end{figure}

\section{Conclusion}

Simulation of electron transport in short period GaN/AlGaN SLs using a single-electron Monte Carlo approach with a 1-band non-parabolic band structure are presented and show electron transport not greatly reduced from GaN bulk values, at least for energy values below the onset of IV scattering. Compared with bulk GaN, the peak velocity of GaN/Al$_{0.2}$Ga$_{0.8}$N 4MLs/4MLs reduce 8\%, and 14\% for GaN/Al$_{0.2}$\allowbreak Ga$_{0.8}$N 6MLs/6MLs. The energy of the electrons that can be treated within this model neglecting IV scattering, is high enough to show velocity overshoot. 

A detailed electron transport study using a single electron Monte Carlo approach using a 3-band model in bulk GaN, AlN and AlGaN was used to investigate the role of IV scattering for high F field transport. We give a new method for the selection of the 3-band model for AlGaN to improve the analysis of IV scattering. This study was used to aid in developing a 1-band model to use in the Monte Carlo treatment in the SL. The effective mass and non-parabolic factors of the lowest miniband of GaN/Al$_x$Ga$_{1-x}$N SLs with $L_{QW}$=$L_{QB}$ are calculated. The scattering process in the SLs is analyzed and the velocity-field curves of a number of short-period SLs with low Al composition are presented.  It is found that low Al composition in the QB and short L are beneficial in resulting in a low effective mass for the miniband and creating a wide energy miniband and therefore good transport properties for the electrons in vertical transport.

We provide a method for simulating the electron transport properties of GaN/Al$_x$Ga$_{1-x}$N short-period superlattices which is fast, convenient and applicable to a range of devices employing short-period SLs and therefore useful for semiconductor device design. In particular, since hole transport in GaN and AlGaN alloys is poor, hot electron transistors in GaN–based materials, devices that employ only electron transport might be a good solution.

\medskip

%

\appendix
\section{Additional Tables}

\begin{table}[h]
 \centering
  \begin{tabular}[htbp]{@{}lll@{}}
    \hline
    Parameters & GaN & AlN \\
    \hline
    Optical phonon energy (meV)[1]                    &91.2                    &99.2                \\
    Static dielectric constant[1]                    &8.9                     &8.5                 \\
    Acoustic deformation potential (eV)[1]           &8.3                     &9.5                 \\
    High-frequency dielectric constant[1]            &5.35                    &4.77                \\
    Piezoelectric constant, $e_{14}$ (C/${cm}^2$)[1] &3.75$\times{10}^{-5}$   &9.2$\times{10}^{-5}$\\
    Inter-valley deformation potentials (eV/cm)[1]    &${10}^{9}$              &${10}^{9}$          \\
    Inter-valley phonon energies (meV)[1]             &91.2                    &99.2                \\
    Transverse sound velocity (cm/s)[1]              &2.68$\times{10}^{5}$    &3.70$\times{10}^{5}$\\
    Longitudinal sound velocity (cm/s)[1]            &6.56$\times{10}^{5}$    &9.06$\times{10}^{5}$\\
    Strain[31]                                        &0(sapphire)             &2.48\%              \\
    Conduction band offset[33]                        &0                       &1.2                  \\
    Lattice constant(nm)[28]                          &0.26                    &0.25                 \\
    Acceptor ionization energy(meV)[34][35]               &30                      &250                 \\
    Piezoelectric constant(${10}^{-12}$m/V)[31]      &$d_{31}$=-1.7 $d_{15}$=-1.7  $d_{33}$=3.4   &$d_{31}$=-2 $d_{15}$=-2  $d_{33}$=4\\
    Spontaneous polarization constant(C/${m}^2$)[31]  &-0.029                  &-0.081             \\
    Effective mass(x)[38]                                 &0.2$m_0$                &0.3$m_0$            \\
    Effective mass(y)[38]                                 &0.2$m_0$                &0.3$m_0$             \\
    Effective mass(z)[1]                                 &0.2$m_0$                &0.48$m_0$              \\
    \hline
  \end{tabular}
   \caption{The constant parameter corresponds to bulk wurtzite GaN, AlN}
   \label{tab:appendix_table}
\end{table}

\textbf{References}\\
1   S. K. O'leary, B. E. Foutz, M. S. Shur, L. F. Eastman, J. Mater. Sci.: Mater. Electron. 2006, 17, 87.\\
2   M. Razeghi, IEEE Photonics J. 2011, 3, 263.\\
3   F. Chen, X. Ji, S. P. Lau, Mater. Sci. Eng. R: Rep. 2020, 142, 100578.\\
4   Y. Zhang, J. Singh, J. Appl. Phys. 1999, 85, 587.\\
5   A. P. Zhang, L. B. Rowland, E. B. Kaminsky, V. Tilak, J. C. Grande, J. Teetsov, A. Vertiatchikh, L. F. Eastman, J. Electron. Mater. 2003, 32, 388.\\
6   A. El-Ela, A. Z. Mohamed, Int. Scholarly Res. Notices 2013.\\
7   J. D. Albrecht, R. P. Wang, P. P. Ruden, M. Farahmand, K. F. Brennan, J. Appl. Phys. 1998, 83, 1446.\\
8   J. Fang, M. V. Fischetti, R. D. Schrimpf, R. A. Reed, E. Bellotti, S. T. Pantelides, Phys. Rev. Appl. 2019, 11, 044045.\\
9   F. Bertazzi, M. Moresco, E. Bellotti, J. Appl. Phys. 2009, 106, 063718.\\
10  S. Yamakawa, R. Akis, N. Faralli, M. Saraniti, S. M. Goodnick, J. Phys.: Condens. Matter 2009, 21, 174206.\\
11  N. Vogiatzis, J. M. Rorison, In Semiconductor Modeling Techniques, Springer Berlin Heidelberg, Berlin, Heidelberg 2012, pp. 115-152.\\
12  H. Sakaki, in Proc. 17th Int. Conf. Phys. Semiconductors: San Francisco, California, USA, Aug. 6–10, 1984, Springer New York, New York, NY 1985.\\
13  A. D. Bykhovski, B. L. Gelmont, M. S. Shur, J. Appl. Phys. 1997, 81, 6332.\\
14  N. Shuji, S. Masayuki, N. Shin–ichi, I. Naruhito, Y. Takao, M. Toshio, K. Hiroyuki, S. Yasunobu, K. Tokuya, U. Hitoshi, S. Masahiko, Jpn. J. Appl. Phys. 1997, 36, L1568.\\
15  J.-Y. Duboz, Semicond. Sci. Technol. 2014, 29, 035017.\\
16  M. Moresco, F. Bertazzi, E. Bellotti, J. Appl. Phys. 2009, 106, 063719.\\
17  S. K. O'Leary, B. E. Foutz, M. S. Shur, U. V. Bhapkar, L. F. Eastman, Solid State Commun. 1998, 105, 621.\\
18  P. Siddiqua, S. K. O’Leary, J. Mater. Sci.: Mater. Electron. 2018, 29, 3511.\\
19  M. Goano, E. Bellotti, E. Ghillino, C. Garetto, G. Ghione, K. F. Brennan, J. Appl. Phys. 2000, 88, 6476.\\
20  M. Farahmand, C. Garetto, E. Bellotti, K. F. Brennan, M. Goano, E. Ghillino, G. Ghione, J. D. Albrecht, P. P. Ruden, IEEE Trans. Electron Devices 2001, 48, 535.\\
21  C. Jacoboni, P. Lugli, The Monte Carlo Method for Semiconductor Device Simulation, Springer Science and Business Media, 2012.\\
22  D. R. Naylor, Doctoral Dissertation, University of Hull, 2012.\\
23  T. Deguchi, D. Ichiryu, K. Toshikawa, K. Sekiguchi, T. Sota, R. Matsuo, T. Azuhata, M. Yamaguchi, T. Yagi, S. Chichibu, S. Nakamura, J. Appl. Phys. 1999, 86, 1860.\\
24  V. Y. Davydov, Y. E. Kitaev, I. N. Goncharuk, A. M. Tsaregorodtsev, A. N. Smirnov, A. O. Lebedev, V. M. Botnaryk, Y. V. Zhilyaev, M. B. Smirnov, A. P. Mirgorodsky, O. K. Semchinova, J. Crystal Growth 1998, 189, 656.\\
25  W. A. Hadi, M. S. Shur, S. K. O’Leary, J. Mater. Sci.: Mater. Electron. 2014, 25, 4675.\\
26  J. G. Ruch, W. Fawcett, J. Appl. Phys. 1970, 41, 3843.\\
27  B. E. Foutz, et al., Appl. Phys. Lett. 1997, 70, 2849.\\
28  W. Sun, C.-K. Tan, N. Tansu, Sci. Reports 2017, 7, 11826.\\
29  P. Harrison, A. Valavanis, Quantum Wells, Wires and Dots: Theoretical and Computational Physics of Semiconductor Nanostructures, 2016.\\
30  R. E. Bank, R. K. Smith, in Thirteenth Int. Symp. Domain Decomposition Methods for Partial Differential Equations, Domain Decomposition Press, Bergen, 2001, pp. 15-26.\\
31  S.-H. Park, Jpn. J. Appl. Phys. 2000, 39, 3478.\\
32  S.-H. Park, S.-L. Chuang, J. Appl. Phys. 2000, 87, 353.\\
33  Eberlein, T. A. G., Appl. Phys. Lett. 2007, 91, 13.\\
34  P. Pampili, P. J. Parbrook, Mater. Sci. Semicond. Process. 2017, 62, 180.\\
35  M. L. Nakarmi, et al., Appl. Phys. Lett. 2004, 85, 3769.\\
36  Ridley, B.K., Quantum processes in semiconductors, Oxford University Press, USA, 2013.\\
37  C.-Z. Zhao, T. Wei, L.-Y. Chen, S.-S. Wang, J. Wang, Superlattices Microstruct. 2017, 109, 758.\\
38  Vurgaftman, I., Meyer, J.R., J. Appl. Phys. 2003, 94, 3675-3696.\\

\section*{Acknowledgements}
The authors would like to express their gratitude to the developers of nextnano software for providing a valuable tool in the simulation of our research results.

\section*{Author contributions statement}
Mengxun and Judy contributed to this work. Mengxun wrote the article as the first author. Both authors conceived and designed the simulation. Mengxun conducted the model, analyzed the data and performed the numerical simulations. Judy developed the theoretical framework as a supervisor.




\end{document}